\documentclass[a4]{article}
\usepackage[utf8]{inputenc}
\usepackage{authblk}

\usepackage{setspace}

\usepackage{lineno}
\usepackage{graphicx}
\graphicspath{{Figures/}}
\usepackage{amsmath}
\usepackage{amssymb}
\usepackage{pifont}
\usepackage{bm}
\usepackage[semicolon]{natbib}
\usepackage{subcaption}
\usepackage{lscape}
\setcitestyle{aysep={},citesep={,}}
\usepackage{tikz}
\usetikzlibrary{shapes,arrows}

\usepackage{titlesec}
\titleformat*{\subsubsection}{\normalfont}

\setcitestyle{notesep={; },round,aysep={},yysep={;}} 

\title{Spatial and temporal evolution of an experimental debris flow, exhibiting coupled fluid and particulate phases}
\author[1]{C. M. Chalk}
\affil[1]{EPSRC CDT in Fluid Dynamics, University of Leeds, Leeds LS2 9JT, United Kingdom} 
\author[2]{J. Peakall}%
\author[2]{G. Keevil}
\affil[2]{School of Earth and Environment, University of Leeds, Leeds LS2 9JT, United Kingdom}
\author[3]{R. Fuentes}
\affil[3]{School of Civil Engineering, University of Leeds, LS2 9JT, United Kingdom}
\affil[3]{Escuela de Ingenieros de Caminos, Canales y Puertos, Universitat Polit{\`e}cnica de Val{\`e}ncia, Spain}
\date{}

\begin{document}
\maketitle

\newpage

\begin{abstract}

The internal behaviour of debris flows provides fundamental insight into the mechanics responsible for their motion. We provide velocity data within a small-scale experimental debris flow, consisting of the instantaneous release of a water-granular mixture along a rectangular flume, inclined at $31^{\circ}$. The results show a transition from a collisional, turbulent front to a viscous-type, steady flow body, exhibiting strong fluid-particulate coupling. This is the first time that both the spatial and temporal evolution of the internal mechanics of a small-scale debris flow have been considered. Our results serve as invaluable data for testing two-phase fluid-particulate numerical models.

% By comparing with the results from experimental results in the literaure, our results  

%\keywords{Debris flow, Particle Image Velocimetry, Flume experiment, Multi-phase flow}

\end{abstract}

%\linenumbers

\section{Introduction}

%\large

A debris flow is a gravity-driven flow where the interaction of both solid and fluid phases governs the dynamics \citep{iverson1997physics}. Debris flows exhibit extremely complex and destructive behaviour \citep{costa1984physical}, and a significant amount of research has been dedicated to understanding the governing physical processes (e.g. \cite{takahashi1981debris,iverson1997physics,kaitna2007experimental,iverson2014depth,pastor2018two}).By nature, the occurrence of debris flows is unpredictable and their behaviour is dependent on a number of physical conditions (such as terrain and material composition). Repeatable, physical models are able to capture the key features of real debris flows, allowing the investigation of flow dynamics in a controlled environment. Furthermore, experimental debris flows are invaluable for the validation of mathematical and numerical models.

Many different experimental set-ups have been employed to investigate one or more aspects of debris flow behaviour, within a wide range of physical scales \citep{takahashi1978mechanical,gregoretti2000initiation,armanini2005rheological,larcher2007set,iverson2010perfect,johnson2012grain,de2015effects}. Large scale experiments have the benefit of being directly comparable to real debris flows \citep{iverson2015scaling}, yet they are expensive, time consuming and complicated to execute. Alternatively, small scale experiments have the advantage of being simple and repeatable, and are capable of reproducing real debris flow features (such as the formation of a distinct granular `head' and fluid-like `body')  \citep{paleo2014fluid,turnbull2015debris,lanzoni2017coarse}. Perhaps of most significance, experimental debris flows at a small scale allow for the observation of internal flow dynamics --- enabling the calculation of internal velocity profiles, thus revealing the mechanical and rheological behaviour of the fluid-granular mixture. 

%Such information is essential for an informed prediction of debris flow behaviour, as well as the development of numerical models.

Various small scale experiments have been dedicated to the analysis of the internal velocity profiles within debris flows --- typically performed in transparent flumes, with cameras recording the flow \citep{armanini2005rheological,kaitna2014surface,lanzoni2017coarse,sanvitale2016visualization}. Image processing techniques are applied to the flow images to calculate material displacement and obtain internal velocity profiles. \citet{armanini2005rheological} examined the velocity profiles within a series of experiments consisting of a recirculating mixture of polyvinyl chloride (PVC) pellets and water in an inclined flume, over an erodible and non-erodible bed. The distinct shapes of the profiles revealed four different granular flow regimes --- immature, mature, plug flow and solid bed flow. These definitions have been frequently used to classify the results of experimental debris flows conducted in more recent years \citep{kaitna2014surface,lanzoni2017coarse,sanvitale2016visualization}. Vertical velocity profiles can also be compared directly with analytical profiles --- based on simplified mathematical models with an assumed rheology \citep{bagnoldexperiments} --- which may exhibit granular or viscous-type behaviour. In the experiments conducted by \citet{kaitna2014surface} and \cite{sanvitale2016visualization}, the type of velocity profile within debris flow bodies was found to be dependent on the mixture composition. Alternatively, one single rheological model may be insufficient to describe the velocity profile throughout the whole flow depth. \citet{lanzoni2017coarse} analysed the internal velocity profiles in a series of flume experiments consisting of a steady debris flow over an erodible bed. The profiles showed a gradual spatial transition from frictional behaviour near the bed (dominated by lubricated, frictional grain contacts), to inertial in the upper half of the flow (dominated by granular collisions). 

Transient, collisional flow behaviour can be identified by considering velocity deviations from the mean, over a specified time period. This approach was taken by \citet{paleo2014fluid} in the analysis of a series of dam break experiments of fluid and glass sphere mixtures. A specified velocity threshold was used to establish the areas of the flow dominated by granular collisions and fluid turbulence --- distinguished by a high velocity deviation over the time-averaged period. The remaining areas of the flow exhibited \textit{non-fluctuating} behaviour. \citet{paleo2014fluid} employed this technique in order to remove collisional, unsteady data from their analysis. However, the method provides a useful and relatively simple way to divide debris flow behaviour into two broad regimes --- collisional and non-fluctuating.

Although there are several investigations of the internal behaviour and velocity profiles within experimental debris flows \citep{kaitna2014surface,paleo2014fluid,kaitna2016effects,lanzoni2017coarse,sanvitale2016visualization}, we lack information on the internal evolution of such flows. Previous work has only considered the internal behaviour of steady flows, mainly concerning the velocity profiles within the debris flow bodies. This is due to the fact that transient velocity data are subject to error in terms of repeatability, and the vertical profiles cannot be compared with simple mathematical models (which are derived under the assumption of a steady state). However, the temporal evolution of coupled fluid-particulate flows could provide valuable insight into the formation and mechanics of the debris flow head-body architecture. Therefore, we analyse the internal behaviour of a small-scale rapid debris flow, for the entire flow duration. We aim to provide a description of the spatial and temporal internal flow evolution, for fixed experimental conditions. The experiments consist of the dam break release of a water-granular mixture along an inclined flume. We are interested in the internal evolution of a rapid flow that travels along the entire length of the flume (with minimal material deposition), corresponding to the solid bed flow of the regimes defined by \cite{armanini2005rheological}. This represents an extreme case of debris flow propagation in terms of material velocity, at a small scale. We use Particle Image Velocimetry (PIV) to obtain the internal velocity profiles, providing high quality, robust and repeatable data. Data of this quality are rare within the literature, and are ideal for the validation and development of numerical models of debris flows. We utilise the PIV data to infer information on the evolving flow behaviour.

The remainder of this paper is structured as follows. The experimental methodology is detailed in Section \ref{sec:exp_method}, including a description of the PIV method. The velocity data are presented in Section \ref{sec:exp_results}, in the form of flow fields and vertical profiles. We consider the deviations of the velocity, in order to identify collisional and non-fluctuating regimes. The implications of the findings are discussed in detail in Section \ref{sec:exp_discussion}, and the results are compared to those of experimental debris flows within the literature. The key findings of this investigation are summarised in Section \ref{sec:conc}.

\section{Experimental methodology}
\label{sec:exp_method}

\begin{figure}
\centering
\includegraphics[]{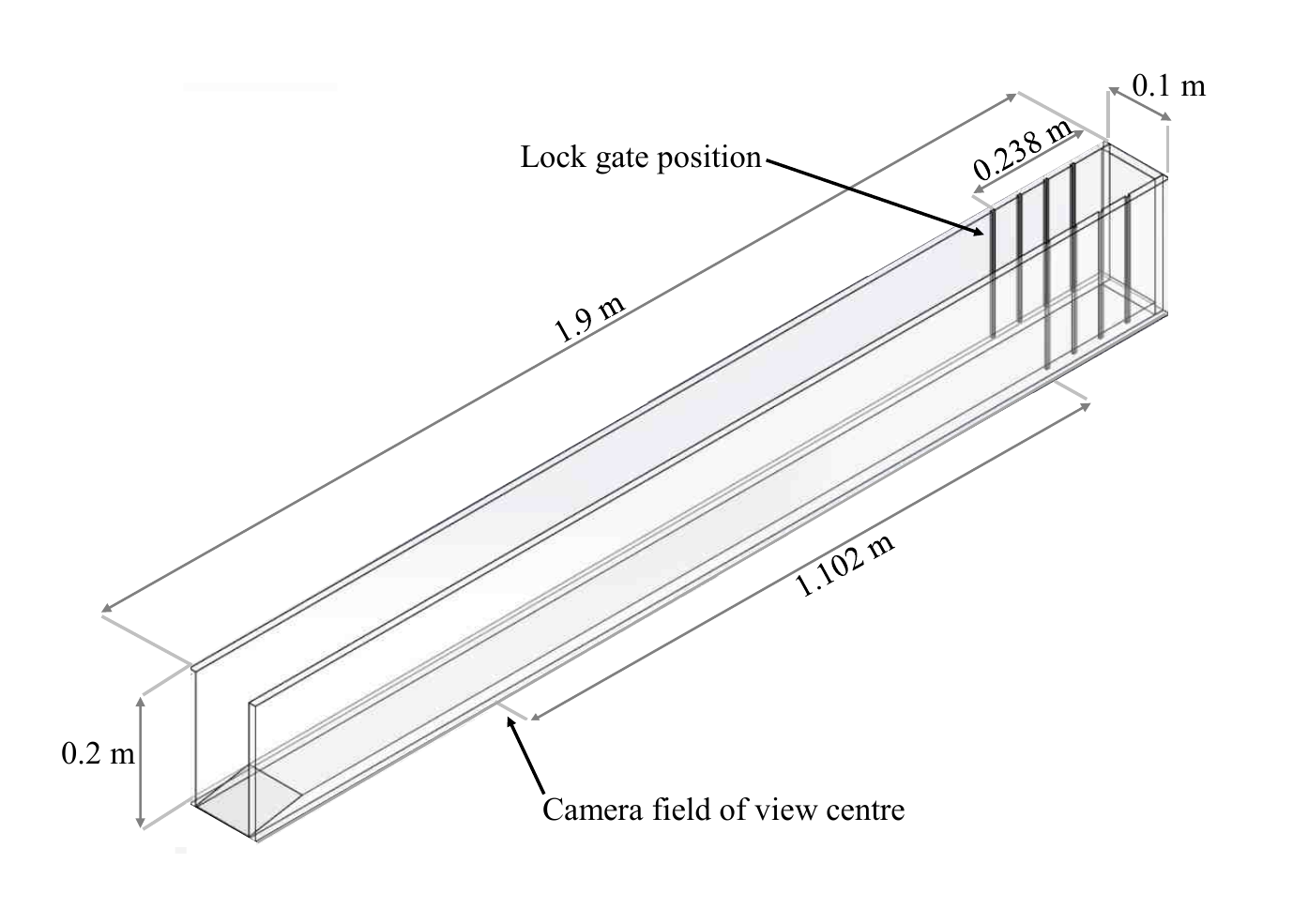}
\caption{A schematic depiction of the experimental flume.}
\label{fig1}
\end{figure}

\begin{figure}
\centering
\includegraphics[width=0.6\textwidth]{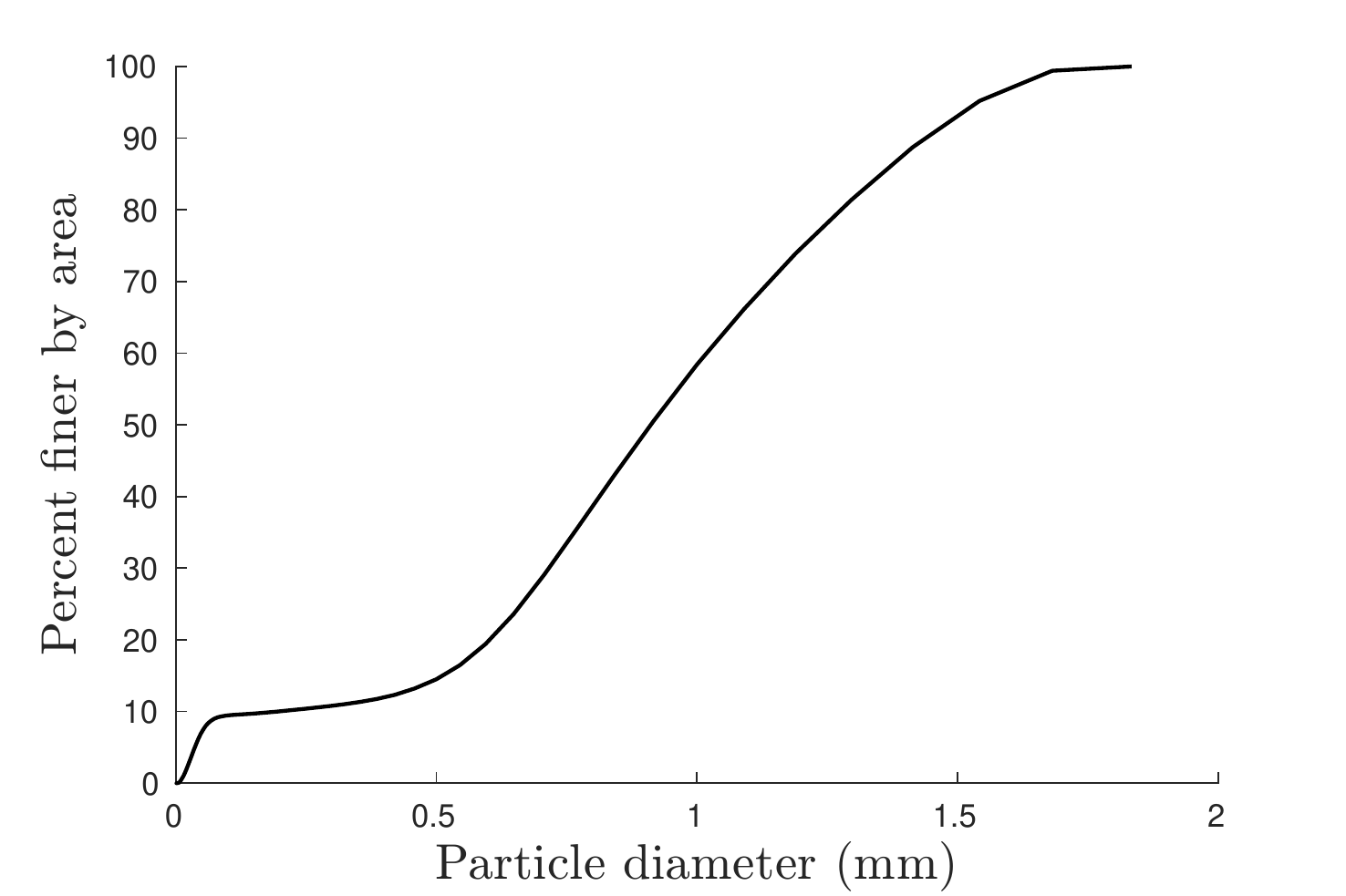}
\caption{Particle size distribution for the glass grit.}
\label{fig2}
\end{figure}

A mixture of water and sediment was manually released from behind a lock gate in a rectangular flume of dimension $1.9 \times 0.2 \times 0.1$ m, at an inclination of $31^{\circ}$ (see Figure \ref{fig1}). This angle of inclination corresponds to that of large scale flume experiments at the United States Geological Survey debris flow flume \citep{iverson2010perfect}, and enables a rapid flow propagation. The mixture consisted of 2.177 kg of sediment and 1.5 l of water, resulting in a total volume of 0.0026775 m$^3$, with an initial solid volume fraction of $\phi_s = 0.44$. The sediment was composed of multicoloured, crushed glass grit with an angular shape, to represent natural granular material. The particle size distribution is shown in Figure \ref{fig2}. The mean particle size is $d_{50} = 0.917$ mm, where $d_x$ denotes the percentage passing by area. The coefficient of uniformity $C_U = d_{60}/d_{10}$ represents the particle size variety, where $d_{60} = 1$ mm, $d_{10} = 0.1928$ mm and $C_U = 5$ (to the nearest integer). The finer particles are expected to contribute to the viscous effects that are frequently observed in granular flows \citep{iverson1997physics}. Sediment of the same grade was permanently fixed onto the flume bed to generate roughness which would produce a no-slip flow. Due to the high friction created by the bed roughness, we found that mixtures with a volume fraction less than $\phi_s = 0.44$ did not propagate along the length of the flume (and were therefore not representative of solid bed flows).

\begin{figure}
\centering
\includegraphics[]{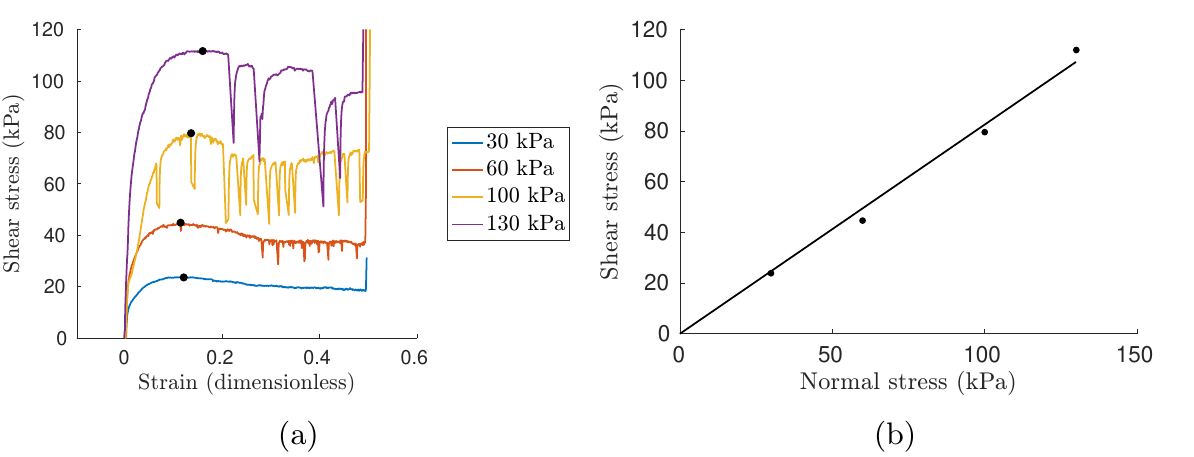}
\caption{Shear box test data: a) Stress-strain relationship, b)Yield surface.}
\label{fig3}
\end{figure}

A shear box test was conducted to determine the mechanical properties of the granular material. The data obtained for the saturated glass grit are provided in Figure \ref{fig3}, for normal stress values of 30 kPa, 60 kPa, 100 kPa and 130 kPa. The samples were inspected after completion of the tests and no particle crushing was observed. A linear fit is applied to the relationship between the normal stress and the peak shear stress, as shown in Figure \ref{fig3}b. The gradient and the $y-$intercept of this fit correspond to the internal friction angle and the cohesion of the material respectively. The glass grit was found to be non-cohesive, with an internal friction angle of $39^{\circ}$. The shear modulus of the material can be approximated as the gradient of the strain-stress curve before the peak values. This was found to be approximately $2.66 \times 10^5$ Pa.

%See \citet{powrie2004} for further details on shear box tests.

\begin{figure}
\centering
\includegraphics[width=0.8\textwidth]{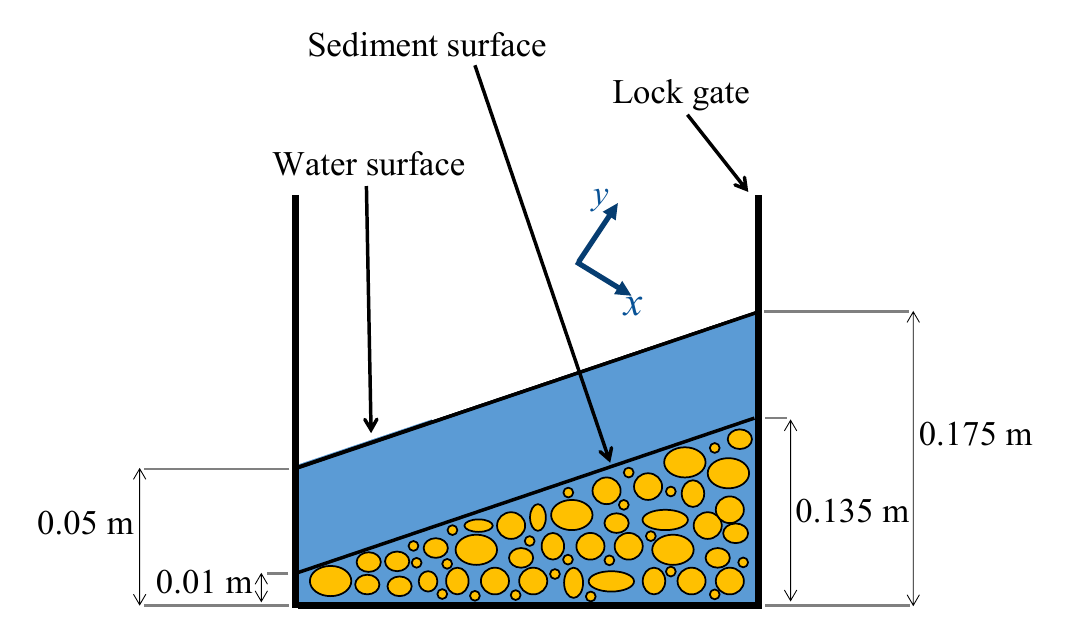}
\caption{The initial placement of the water-granular mixture behind the lock gate.}
\label{fig4}
\end{figure}

At the beginning of each experimental run, 2.177 kg of glass grit was placed behind a lock gate with a cross-sectional area in the shape of a trapezoid, occupying a volume of $0.0017255$ m$^3$. Subsequently, 1.5 l of water was added slowly to minimise the disturbance to the top of the sediment. Due to its porosity, the sediment was rapidly saturated fully. The initial placement of the sediment and water is depicted in Figure \ref{fig4}, where the bottom layer consists of a mixture of water and glass grit, while the top layer is composed of water only.

\begin{figure}
\centering
\includegraphics[width=0.6\textwidth]{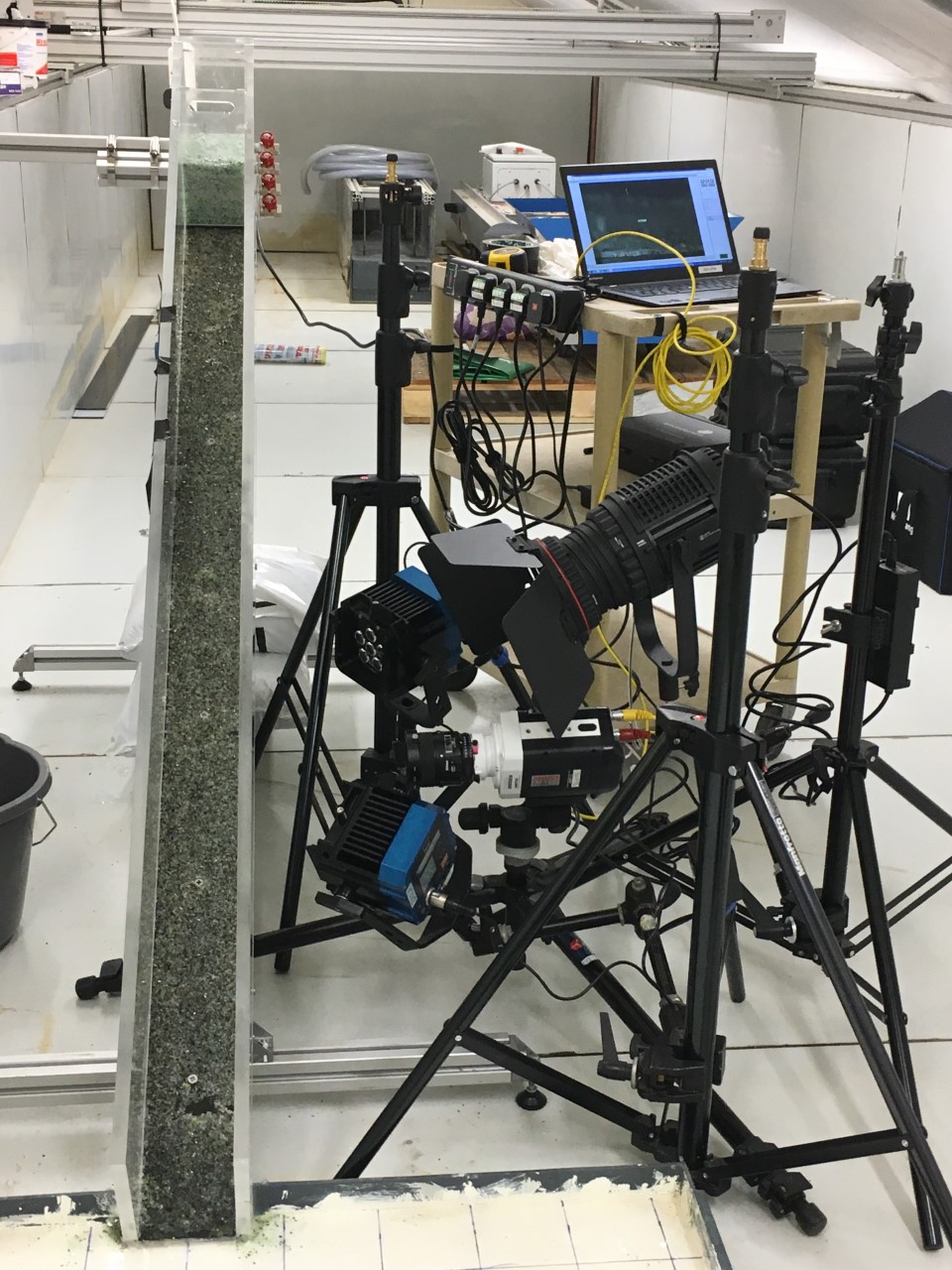}
\caption{A photograph of the experimental set-up just prior to flow initiation. The set-up consists of an acrylic channel with a roughened bed, that runs out onto a broader surface (at the bottom of the picture). A high speed camera and multiple LED lamps are used to visualise the flow.}
\label{fig5}
\end{figure}

To check for repeatability, the debris flow experiments were performed three times. The surface of the sediment phase was marked onto the flume, to ensure that it was placed in the same initial position for each experimental run. To observe the propagation of the flow, a high speed camera was positioned with its centre 1.102 m downstream from the lock gate, with the front of the lens 0.19 m from the flume. The camera is a Vision Research, Miro M120 Colour, with a Zeiss, 50 mm F1.4 ZF2 Planar lens. Upon release of the lock gate, and coeval triggering of the high speed camera, the mixture propagated downstream along the length of the flume and onto the run-out area. In order to capture the rapid flow dynamics, the images were taken at a rate of 1200 frames per second, with an exposure time of 200 $\mu$s and a resolution of 1280 $\times$ 800 pixels. This short exposure time required the addition of extra lighting to  obtain a suitable image quality. For this, two Nila LED lights (model Zaila) were placed on either side of the camera, and one NanGuang LED light (model CN-60F) was positioned above it. The three lights were directed to optimise the light conditions in front of the camera. A photograph of the experimental set-up is shown in Figure \ref{fig5}. Water was poured along the flume bed before and after each experimental run to ensure the removal of any loose sediment that had stuck to the bed. 

\subsection{\textit{Particle Image Velocimetry}}
\label{sec:piv}

A PIV processing method was applied to the images obtained with the high speed camera. This is an experimental technique used within fluid and soil dynamics, where instantaneous velocity fields are determined by tracking the displacements of individual particles, or groups of particles, within a flow \citep{adrian1991particle,white2003soil,adrian2011particle,pinyol2017novel}. The method (in two dimensions) involves splitting each image frame into a number of interrogation areas, within which the movement of particles is tracked between subsequent frames. The displacement is obtained by estimating the cross-correlation between the particle positions within each interrogation area, where the true displacement of each particle group must be separated from the noise created by particles overlapping between frames. This is achieved by applying statistical correlation methods to the data, to determine the most likely `true' particle displacement. An algorithm is then applied to obtain an estimate of the velocity vector field from the displacement values, where certain features of the camera used to obtain the images are taken into consideration. An extensive description of the PIV method can be found in \citet{adrian2011particle}. 

We processed the frames from the camera with the DynamicStudio image processing software to obtain the velocity vectors. The `Adaptive PIV'  option was utilised within DynamicStudio, which automatically adjusts the interrogation area at each frame according to the local particle densities and velocity gradients. This requires the definition of the minimum and maximum values of interrogation areas, which were defined as $32 \times 32$ and $64 \times 64$ pixels respectively. The reader is referred to the DynamicStudio user manual for further details on the adaptive PIV method \citep{dantecuser}. The PIV method often requires the addition of seeding particles to track the fluid movement within each interrogation area of the flow images. This is not typically necessary when considering granular flows, as individual particles are easily detected \citep{white2003soil}. The granular material in the current application is multicoloured --- further facilitating the detection of individual grains --- and no seeding was required.
%Note that the PIV method relies on particle detection, and therefore won't produce accurate results in regions where particles are lacking.  
The PIV analysis was applied to the images of the flow along the side wall, under the assumption of a two-dimensional flow. This is subject to error as the propagating material is unlikely to be uniform across the width of the channel. Furthermore, the flow dynamics are likely to differ somewhat at the flow margin than in the centre, due to the influence of the wall. An alternative option is to use a laser sheet to illuminate a plane in the flow centre, and capture the images in this region for PIV analysis. This method requires the combination of clear particles and a fluid that is refractive index matched, and has been applied recently by \citet{sanvitale2016visualization} to capture the internal dynamics of a granular-fluid mixture in an inclined flume. While the flow dynamics along the flume centre are less influenced by wall effects, this technique does not allow the tracking of dry regions of the flow. 

The rate at which the flow images are captured for analysis with the PIV method must be high enough to capture the local movement of particles. For accurate velocity measurements, the particle displacement between two consecutive frames should not be larger than one quarter of the interrogation area \citep{adrian2011particle}. For this reason, a frame rate of 1200 frames per second was chosen. Time averaged velocity profiles were obtained by averaging the velocity vectors over 30 successive frames, corresponding to a time interval of 0.025 s. The initial flow time ($t = 0$) was defined to be when the front of the flow first reached the field of view of the camera, and the frames were cropped so that the first frame corresponded to $t = 0$. For each considered flow time, the velocities were averaged over the 30 surrounding frames. For example, the velocities at the sixtieth frame ($t = 0.05$ s) were time averaged by averaging over frames 45 to 75. 
     
The velocity data obtained with DynamicStudio are provided in matrix form, where horizontal and vertical velocity components are output in separate matrices at each time frame. The location of the velocity values in each grid correspond to that of the spatial grid, which contains the $x$ and $y$ data. Before post-processing, the matrix dimensions are constant at each frame as the spatial grid does not vary throughout the PIV analysis. For the experimental debris flow, the number of rows of each matrix must be cropped at each frame, to align with the free surface of the flow. This was conducted manually by inspecting the free surface position at each snapshot from the high speed camera. The flow free surface is therefore approximated as a horizontal line.
% In the current work, the post-processing of the PIV data was performed in Matlab.  

\section{Results}
\label{sec:exp_results}

\subsection{\textit{Overall flow behaviour}}

\begin{figure}
\centering
\includegraphics[width=0.65\textwidth]{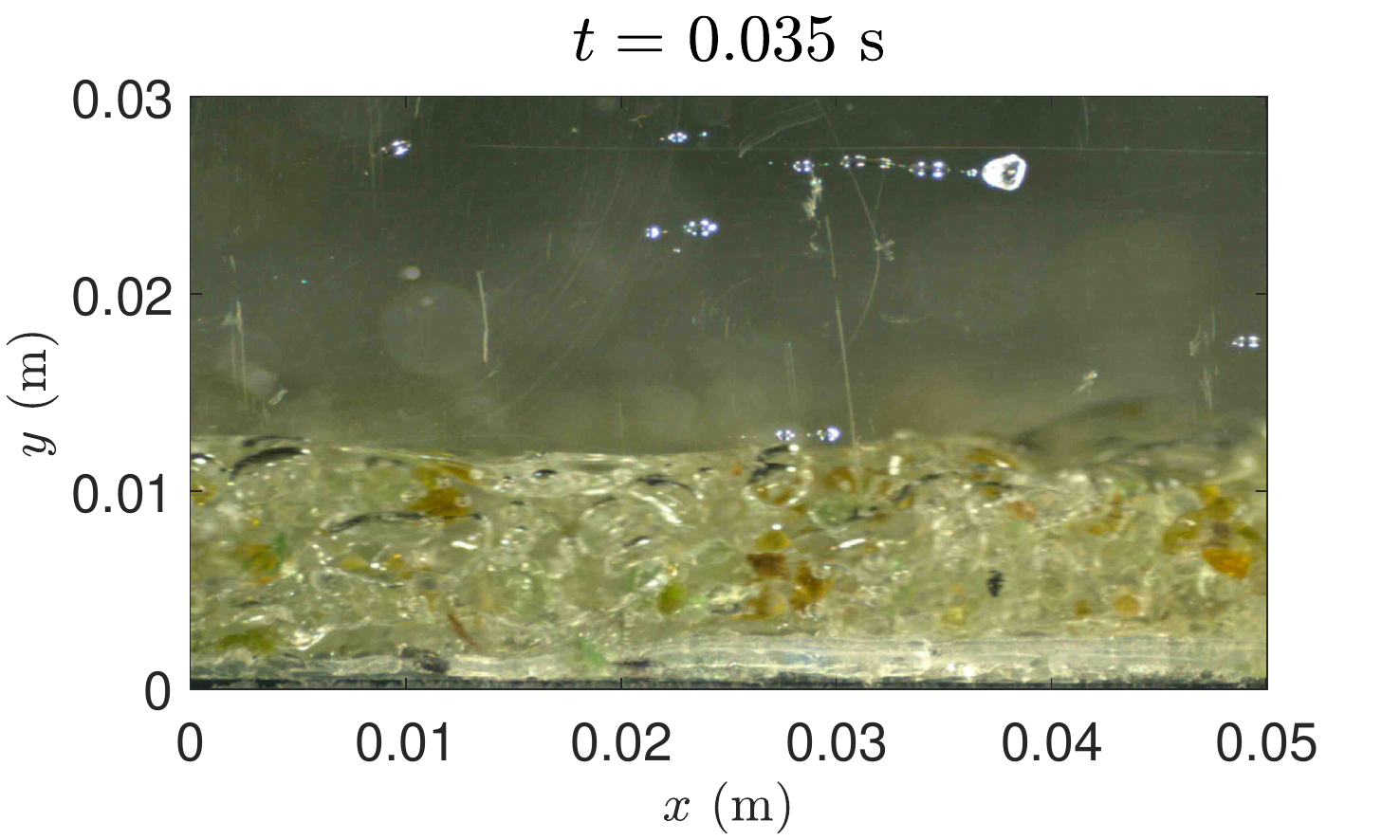}
\caption{A snapshot of the water-granular mixture for Run 1, at $t = 0.035$ s after reaching the field of view of the camera.}
\label{fig6}
\end{figure}

Once released from the lock gate, the water-granular mixture rapidly propagated downstream, reaching maximum front velocities in the range of $1 - 1.2$ m s$^{-1}$. The main bulk of the flow deposited onto the run-out area, although a thin layer of the granular material was deposited along the bed of the flume. The granular material was fully saturated throughout the flow for all repeats of the same experimental run (Run 1, Run 2 and Run 3). A snapshot from the high speed camera at 0.035 s after the material reached the field of view is shown in Figure \ref{fig6}, showing the area that was recorded throughout the flow (0.05 $\times$ 0.03 m$^2$).

\begin{figure}
\centering
\includegraphics{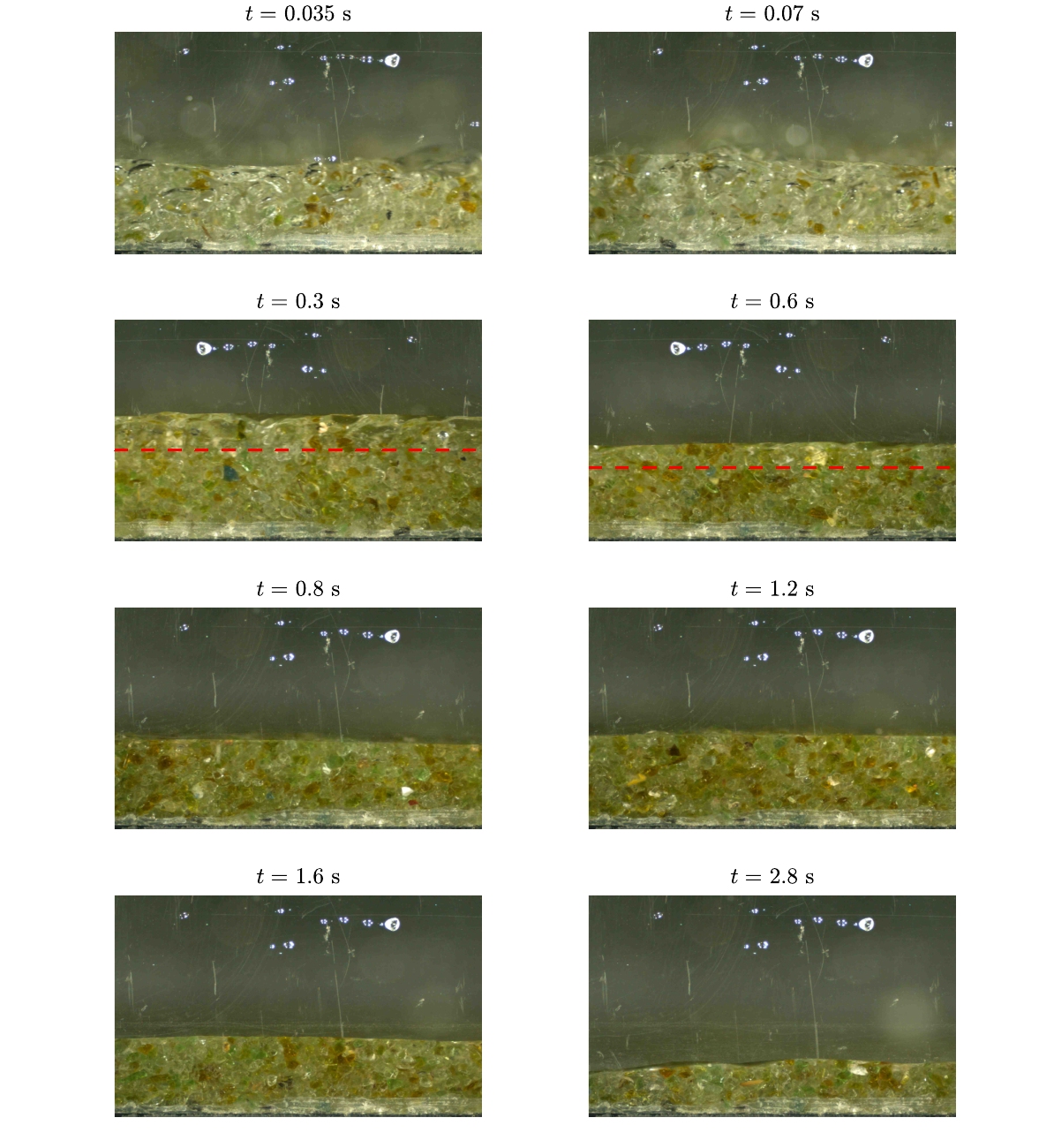}%
\caption{Snapshots of the propagating water-granular mixture for the experimental flow, Run 1. The red, dashed line corresponds to the distinction between a lower, high concentration region and an upper, dilute region (as determined qualitatively). The area of the camera field of view is 0.05 $\times$ 0.03 m$^2$.}
\label{fig7}
\end{figure}

Snapshots of the propagating mixture captured with the high speed camera are provided in Figure \ref{fig7}, for one of the experimental runs. The front part of the flow consists of a dilute and turbulent mixture, with a large void ratio. This region is shown at times $t = 0.035$ s and $t = 0.07$ s in Figure \ref{fig7}. Following this, the height of the flow increases and consists of two visible layers of material, which can be seen clearly at $t = 0.3$ s. The bottom layer is composed of what appears to be a uniform water-granular mixture, while the main constituent of the upper layer is water, along with entrained grains with a large void ratio. The distinction of the two separate layers diminishes as the flow progresses, and the material continues to propagate as a uniform mixture with a constant height. After approximately 1.4 seconds, the flow gradually decreases in height as the material velocity decreases. The mixture comes to a complete rest after 3 seconds, leaving a deposit approximately 0.5 mm in height along the flume. 

%collisional forces in the upper layer, frictional in the bottom  

\subsection{\textit{Flow characterisation}}

The flow behaviour can be characterised by considering the standard deviation $\bar{e}$ of the velocity from the local average, within that interval:
\begin{equation}
\bar{e} = \sqrt{\frac{\sum_{i=1}^N(u_x' - \bar{u}_x)^2}{N-1}},
\end{equation}
where $\bar{u}_x$ is the average velocity over $N$ frames (calculated at a single point), and $u_x'$ is the instantaneous velocity. Low values of standard deviation equate to small variations in instantaneous velocity from the local mean, indicating non-fluctuating behaviour. Conversely, a high standard deviation demonstrates that the averaged velocity profile is not representative of the overall behaviour within the interval, as the flow is rapidly changing. This corresponds to collisional behaviour, which is dominated by fluid turbulence and particle collisions \citep{bagnoldexperiments,johnson1987frictional}. In the small scale debris flow experiments conducted by \citet{paleo2014fluid}, a sufficiently low deviation from the average was defined as being less than $0.15$ m s$^{-1}$. Non-fluctuating behaviour was characterised by a standard deviation below this value, while above this value, the behaviour was collisional. This threshold value was chosen because it clearly distinguished the flow into two distinct regions, which displayed differences in behaviour. 

\begin{figure}
\centering
\includegraphics{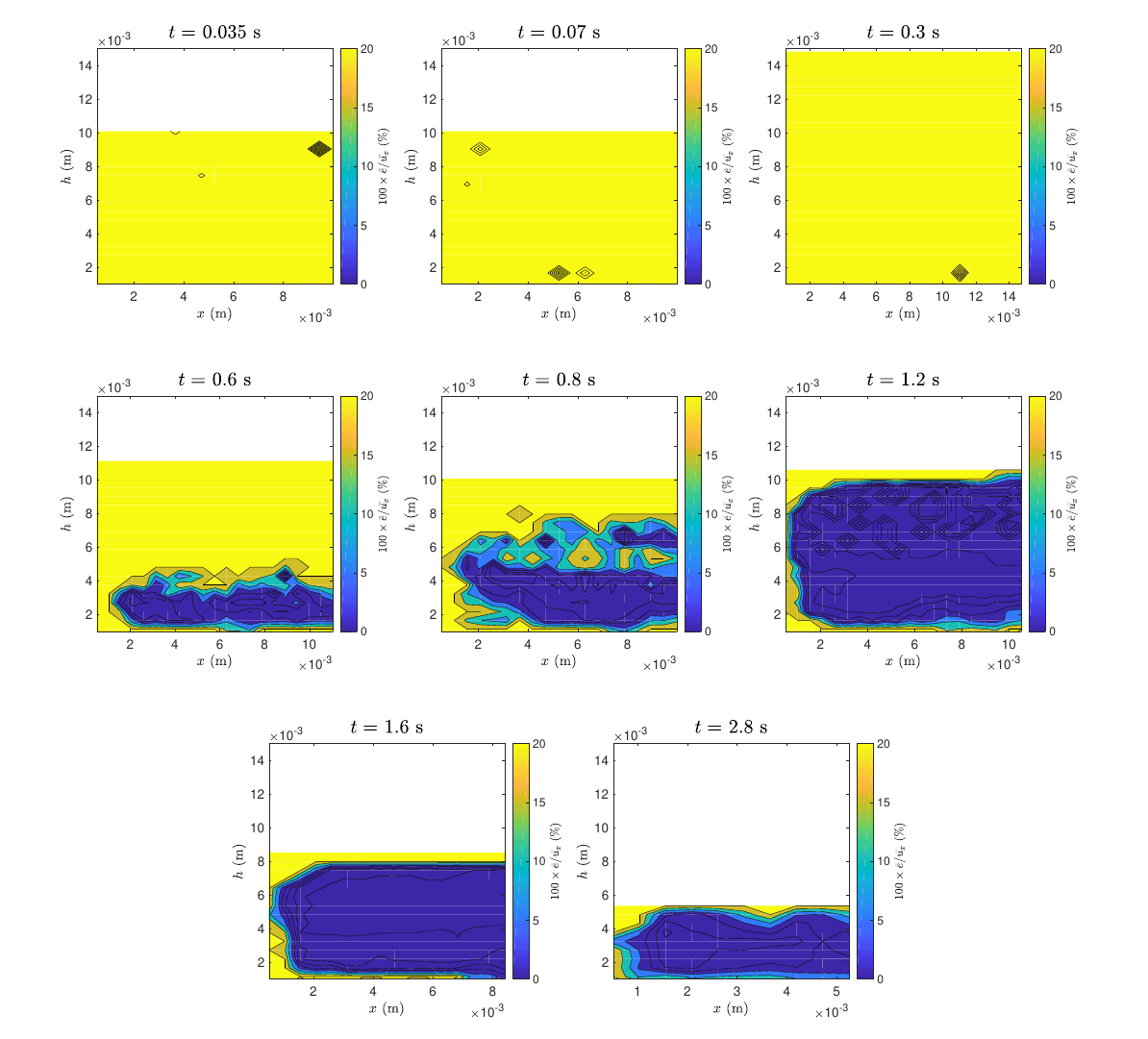}%
\caption{Contour plots of standard deviation as a percentage of the average velocity $100 \times \frac{\bar{e}}{\bar{u}_x}$ at different times of flow for Run 1 of the experiment. The upper limit on the scale bar is defined as $20 \%$ of the velocity average.}
\label{fig8}
\end{figure}

Rather than choose a constant threshold value to distinguish between non-fluctuating and collisional behaviour, we adopt a more general approach and instead consider the deviations of velocity as a percentage of the local average velocity at each time frame. Contour plots of the standard deviation as a percentage are provided in Figure \ref{fig8} for Run 1, which is calculated as $100 \times \frac{\bar{e}}{\bar{u}_x}$. The upper limit of the contour scale is defined as $20 \%$, which has been chosen as the cut-off value to differentiate between the two types of behaviour. The yellow regions in Figure \ref{fig8} correspond to areas of the flow that have a standard deviation that is greater than $20 \%$ of the local time-averaged velocity, and are assumed to be collisional. The blue areas in Figure \ref{fig8} represent a flow with a velocity deviation that is less than $20 \%$ of the average velocity, and can be assumed to be non-fluctuating. A threshold percentage of $20 \%$ was chosen because it clearly separates the flow into the two regions that can be seen in the flow snapshots in Figure \ref{fig7}. The results in Figure \ref{fig8} show that collisional behaviour is exhibited throughout the depth of the flow at $t = 0.035$ s, $t = 0.07$ s and $t = 0.3$ s. At $t = 0.6$ s, a non-fluctuating layer with a thickness of approximately 5 mm has developed. The height of this layer increases with time, and by $t = 1.2$ s the majority of the flow is non-fluctuating. However, the results in Figure \ref{fig8} show that the experimental flow exhibits high deviations with respect to the local average velocity at the free surface and along the left horizontal boundary, for all times presented. The velocity values at the horizontal boundaries are subject to error due to the truncation of the PIV interrogation area. Regarding the flow free surface, a thin, watery layer is present at all times (see Figure \ref{fig7}). The PIV method relies on particle detection over subsequent frames, and therefore produces unreliable results in regions lacking particles. Consequently, the velocity values at the flow free surface are also subject to error. 

\begin{figure}
    \centering
    \includegraphics{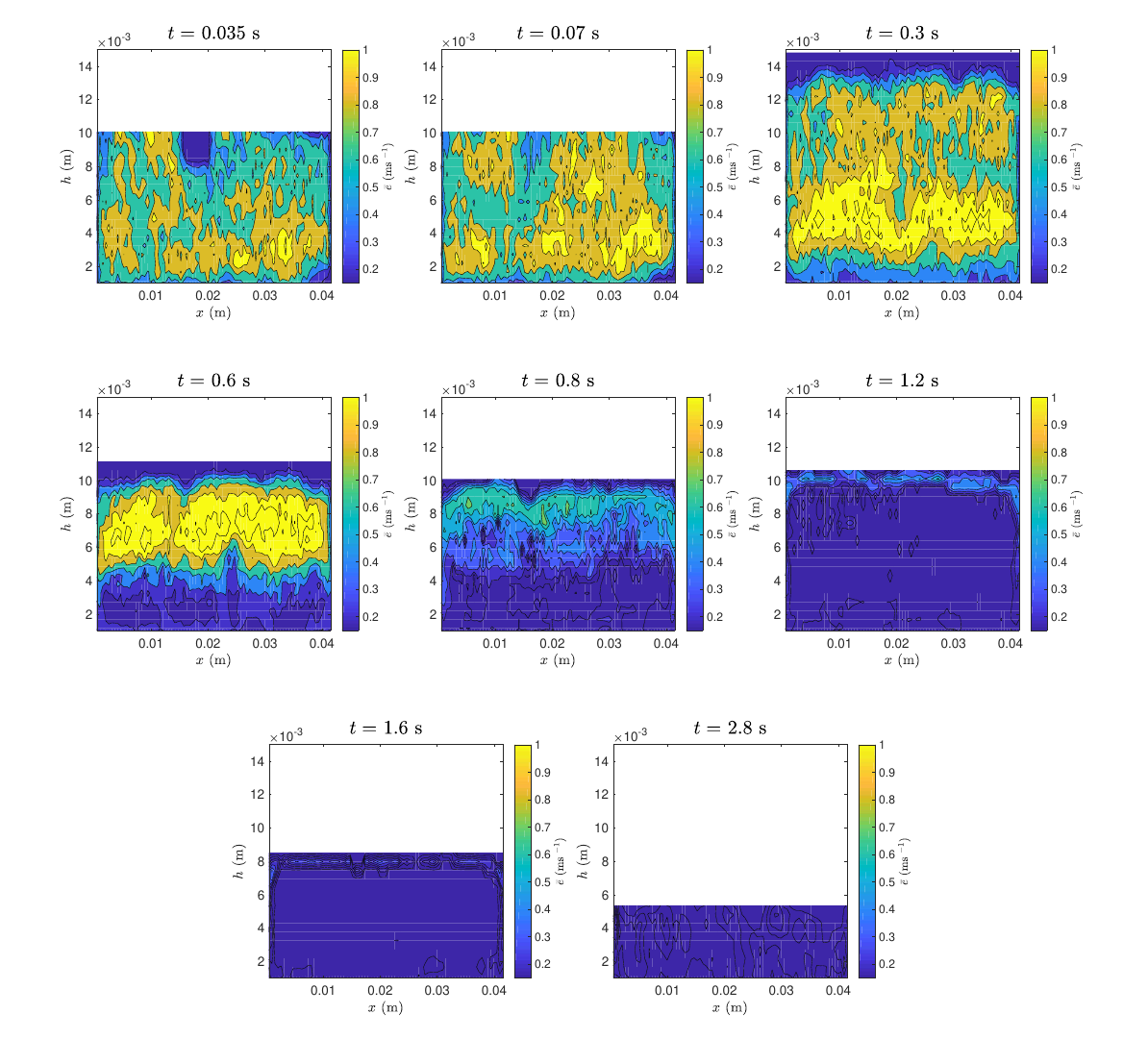}
    \caption{Contour plots of standard deviation $\bar{e}$ at different times of flow for the experimental water-granular mixture, Run 1. The lower limit on the scale bar is defined as $\bar{e} = 0.15$ m s$^{-1}$, which is the suggested cut-off between fluctuating and non-fluctuating behaviour used in the experiments of \citet{paleo2014fluid}.}
    \label{fig9}
\end{figure}

To compare with the method of \citet{paleo2014fluid}, we also use a constant value of $\bar{e} = 0.15$ m s$^{-1}$ to distinguish between non-fluctuating and collisional behaviour. Contour plots of the velocity standard deviation for Run 1 of the experimental flow are provided in Figure \ref{fig9}. The plots in Figure \ref{fig9} are scaled so that the lower limit is equal to  $\bar{e} = 0.15$ m s$^{-1}$. Accordingly, the dark blue areas in Figure \ref{fig9} are assumed to correspond to non-fluctuating regions of the flow, while the lighter colours are assumed to represent the collisional regions. As also shown in Figure \ref{fig8}, the flow transforms from being purely collisional to non-fluctuating, with a layered transition at $t = 0.6$ s and $t = 0.8$ s. However, with a constant threshold value of $\bar{e} = 0.15$ m s$^{-1}$, the contours in Figure \ref{fig9} fail to highlight the high velocity deviations from the local average at the flow free surface, from $t = 0.3$ s onwards (which are revealed in Figure \ref{fig8}). 
 
%After the initial surge of purely collisional flow, the region of collisional behaviour decreases in both height and magnitude with time. Furthermore, the vertical location of the collisional area approaches the free surface with time, until the entire flow is non-fluctuating. 

\begin{figure}
\centering
\includegraphics{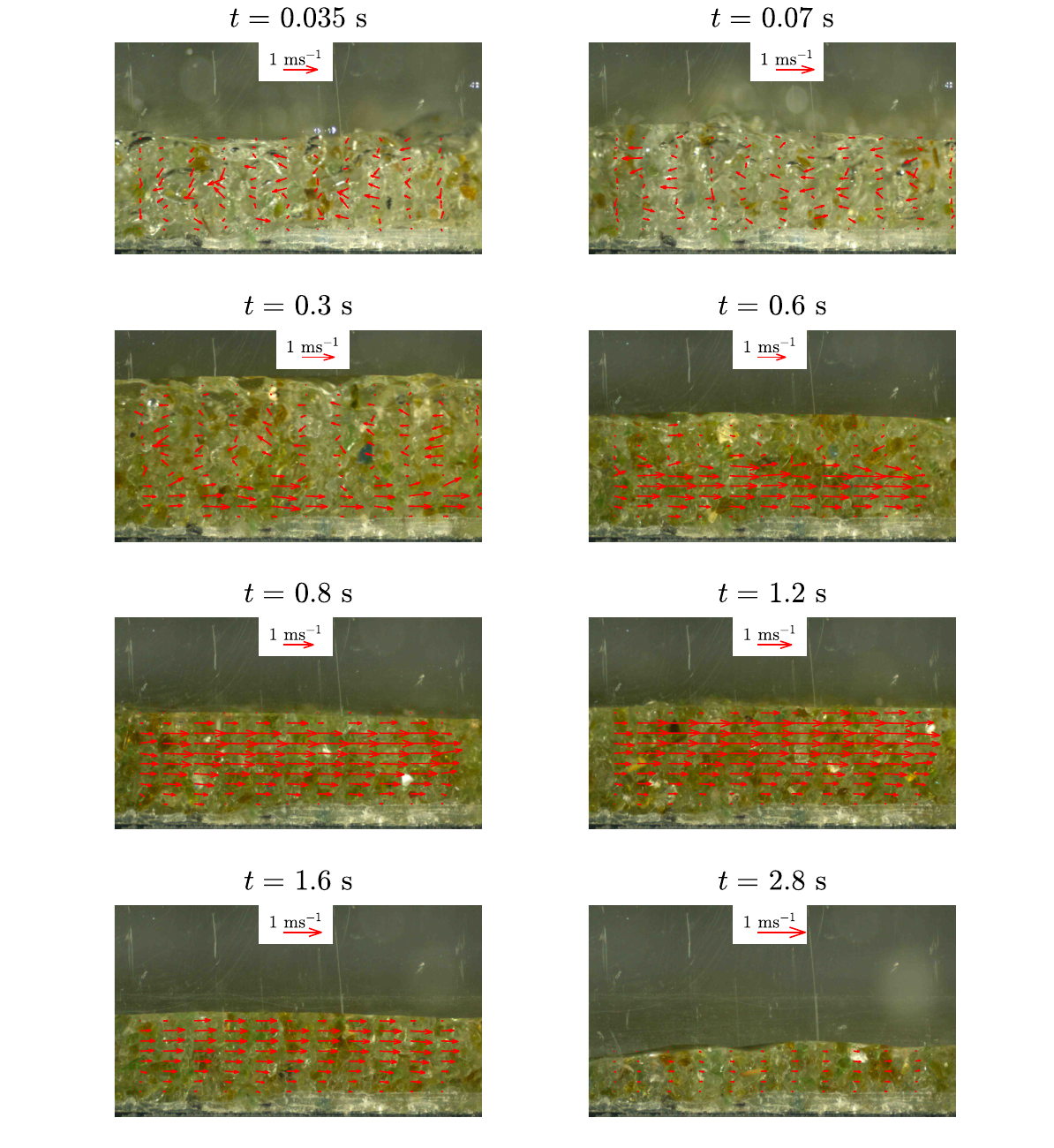}%
\caption{Snapshots of the propagating water-granular mixture for Run 1 of the experimental flow, with the overlaid time-averaged PIV velocity vectors. The area of the camera field of view is 0.05 $\times$ 0.03 m$^2$.}
\label{fig10}
\end{figure}

The averaged velocity vectors obtained from the PIV analysis are superimposed onto the flow snapshots in \ref{fig10} (the vectors are represented by red arrows). The velocity profiles align with the qualitative behaviour of the water-granular mixture described above. At times of $t = 0.035$ s and $t = 0.07$ s, the entire mixture is collisional and the PIV velocity vectors exhibit random and fluctuating behaviour throughout the majority of the flow depth. Most of the vectors are orientated in the negative direction which does not represent the physical behaviour of the flow. The spurious vectors show that there is not a sufficient correlation between particles in successive frames for the PIV method to produce robust velocity vectors. Although these vectors do not represent the true velocity, they indicate the high turbulence of the flow. At $t = 0.3$ s, while the majority of the velocity vectors exhibit similar turbulent behaviour to the earlier flow times, the vectors near the base of the flow are uniformly aligned in the downstream direction. The highest velocities are located in this lower, non-fluctuating region. Subsequently, the height of the non-fluctuating region increases from  $t = 0.3$ s to $t = 0.6$ s, and the velocity vectors correspond to the presence of the two distinct layers discussed above (see Figures \ref{fig8} and \ref{fig9}). The velocity magnitude increases with distance from the bed, and the maximum is located at the top of the non-fluctuating layer. The intensity of the turbulent vector distribution observed between $t = 0.035$ s and $t = 0.6$ s decreases with time. There are a number of negatively orientated vectors in the region between the non-fluctuating and collisional layers at $t = 0.6$ s, indicating the shearing that occurs in this area. At all subsequent times shown, the velocity vectors display linear behaviour throughout the flow depth. Moreover, at $t = 0.8$ s, the two material layers are no longer distinguished, and the height of the velocity maximum is located closer to the free surface. Above the height of the maximum, the velocity decreases towards the flow surface. At $t = 1.2$ s the height of the maximum velocity has increased further, and the same profile is also observed at $t = 1.6$, with a lower overall velocity magnitude. 
By $t = 2.8$ s the velocity has decreased significantly,  and the mixture is almost stationary.

\subsection{\textit{Velocity profiles}}
\label{sec:vel_flow}

\begin{figure}
\centering
\includegraphics{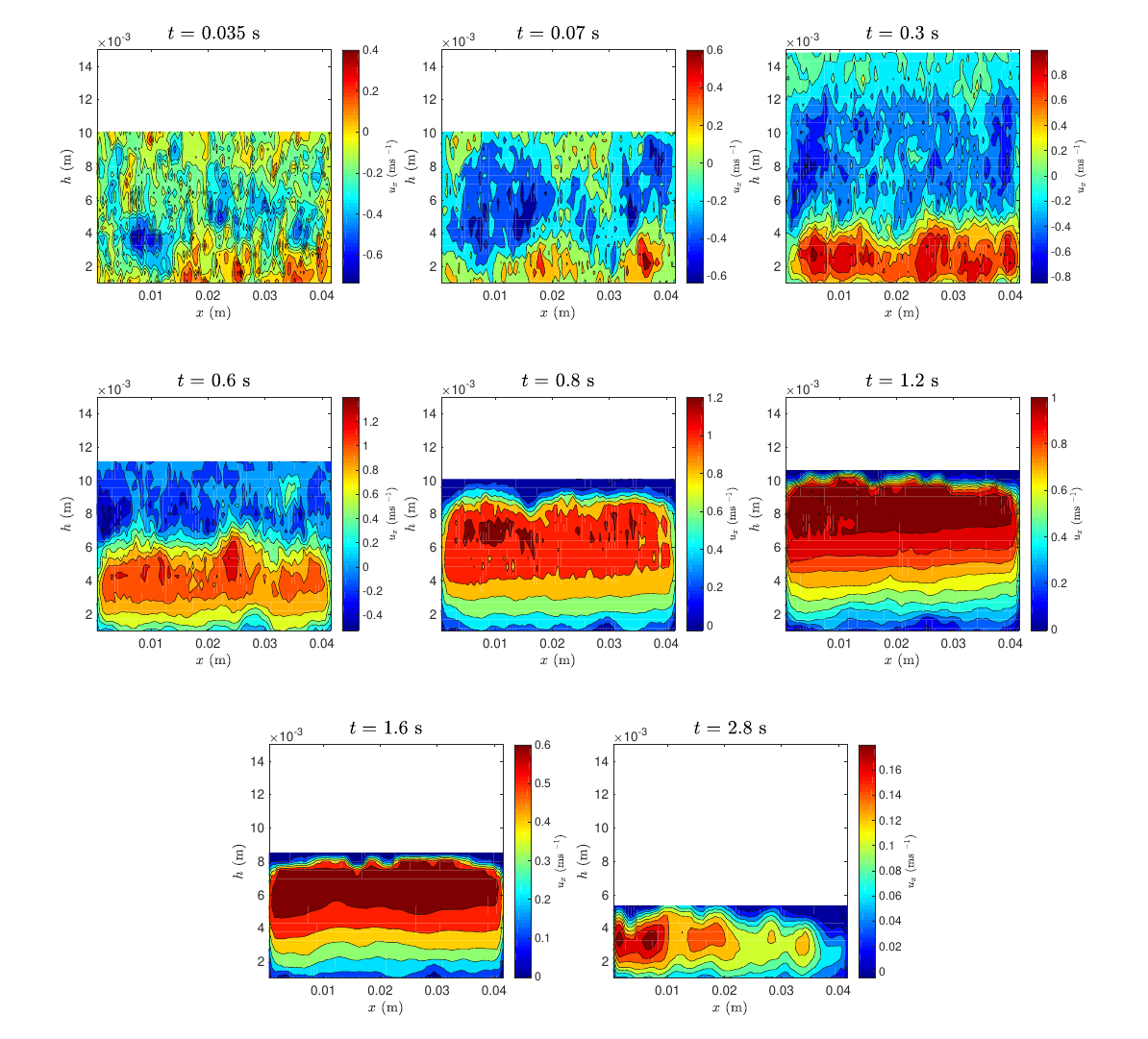}%
\caption{Contour plots of horizontal velocity $u_x$ at different times for Run 1 of the experimental flow. Note the difference in scale for each image.}
\label{fig11}
\end{figure}

The corresponding contour plots for the PIV velocity data are shown in Figure \ref{fig11}. After the initial, fully collisional region of the flow, there is a concentrated area of high velocity at the base at $t = 0.3$ s. Above this, the upper, collisional layer exhibits some negative velocities. As discussed above, these negative velocities are not physical, but are useful in highlighting turbulent and rapidly varying flow regions. Similar behaviour is observed at $t = 0.6$ s, where the height of the concentrated, high velocity region has increased. Between times of $t = 0.8$ s and $t = 1.6$ s, the velocity contours are positive everywhere, and show an increase with height from the flume bed up to a maximum region. Above this region the profiles decrease to approximately zero at the flow free surface, due to the lack of particles detected with the PIV method. In reality, the velocity at the top of the flow is likely to be approximately equal to the velocities in the layer directly beneath it.

\begin{figure}
\centering
\includegraphics{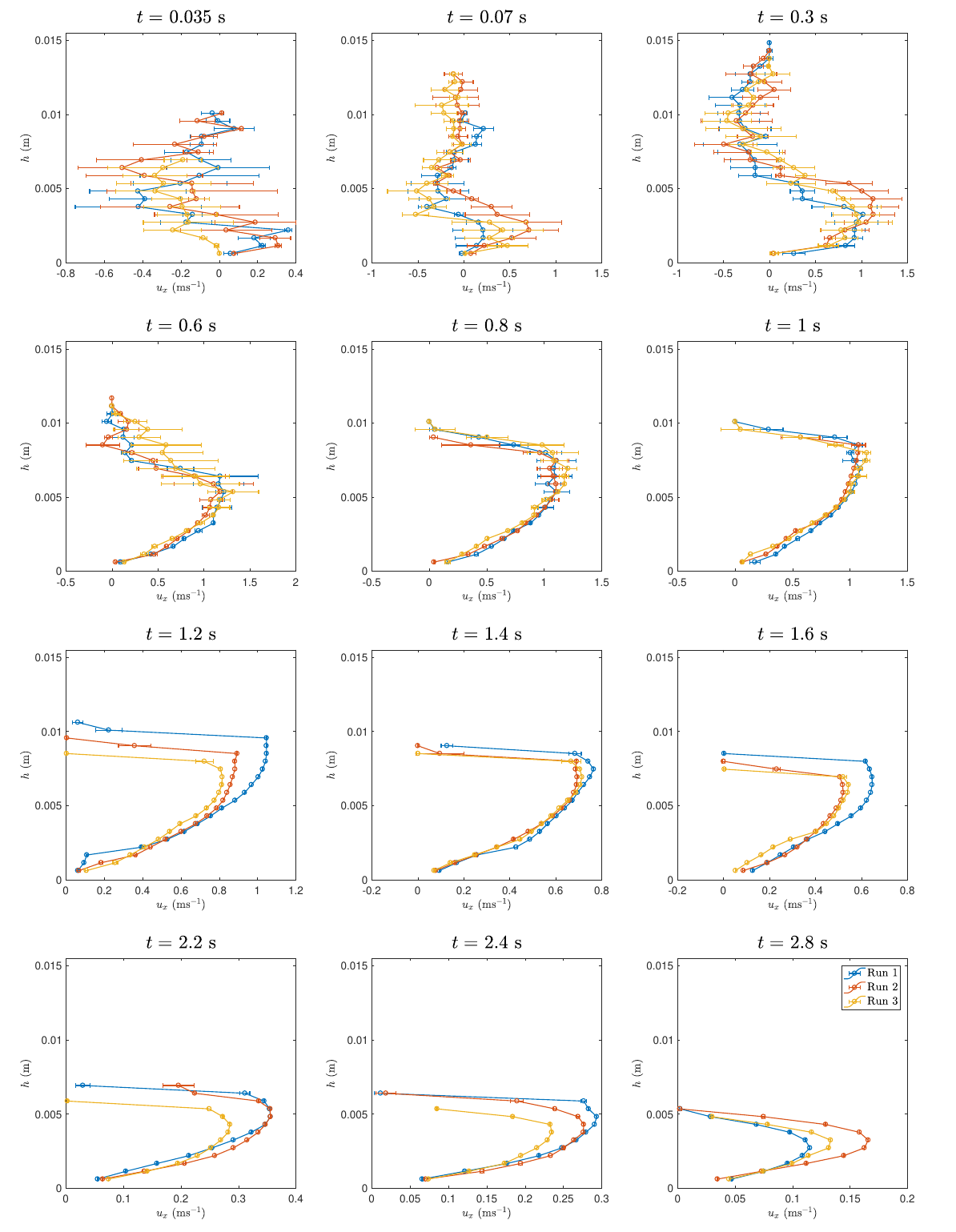}%
\caption{Velocity profiles obtained from the PIV data at various times for the three experimental runs. The horizontal position of the profiles is in the centre of the camera viewing frame (1.102 m downstream from the lock gate).}
\label{fig12}
\end{figure}

Profiles of horizontal velocity $u_x$ against height $h$ are provided in Figure \ref{fig12} for twelve times ranging between $t = 0.035$ s and $t = 2.8$ s, comparing the results from the three different experimental runs. To analyse the error of the time-averaging process, we fitted an autoregressive (AR) model to the instantaneous data over the 30 frames that were averaged. AR models are used to represent a value in a time series as a weighted sum of previous values in the series, and can be used for forecasting purposes \citep{brockwell2002introduction,box2015time}. A time series $X_n$ is defined as a linear combination of past observations $X_{n-1}$ and white noise error terms $\epsilon_n$:
\begin{equation}
X_n = \sum_{i=1}^p \alpha_i X_{n-1} + \epsilon_n,
\end{equation}  
where $\alpha_i$ are the model parameters, and $p$ is the number of past observations that are used in the expression of $X_n$. For the current purpose, the parameter $p$ was assumed to be one, so that each term in the AR model is based on the previous term, and the white noise. The optimum values for the coefficients $\alpha_i$ should minimise the error term. These were determined using the Matlab ARIMA (autoregressive-integrated-moving-average) function, which uses a maximum likelihood method to estimate the model parameters. The error in the experiment can be assessed by considering the residuals between the data and the AR model fit, which are plotted as error bars in Figure \ref{fig12}. Up to $t = 0.3$ s, the error bars are relatively large, indicating that the data cannot be satisfactorily modelled as a first order AR model. The data are unlikely to be stationary in this area, due to the rapidly varying velocities. The error bars are also large in the upper region of the flow at $t = 0.6$ s, where the behaviour is turbulent and collisional. Despite this, the collisional behaviour is consistent between the three different runs, even in the regions that exhibit non-physical negative velocities (between $t = 0.035$ s to $t = 0.6$ s). From $t = 0.8$ s onwards, the residual errors are insignificant. The overall flow behaviour for the three different runs is in good agreement, at each time shown in Figure \ref{fig12} . Small differences in height and velocity can be observed at certain times, namely at $t = 1.2$ s, and from $t = 1.6$ s to $t = 2.8$ s. A number of factors could contribute to such differences, such as the wetness of the bed, a delay in the lock gate release or variable composition between the granular material. Nonetheless, the shape of the velocity profiles align well at all times presented, showing that the experiments are repeatable in terms of the overall flow behaviour. For each run, it can be observed that the height of the velocity maximum steadily increases with time up to $t = 1$ s. From $t = 1.2$ s onward, the location of the velocity maximum decreases with height.

\begin{figure}
\centering
\includegraphics{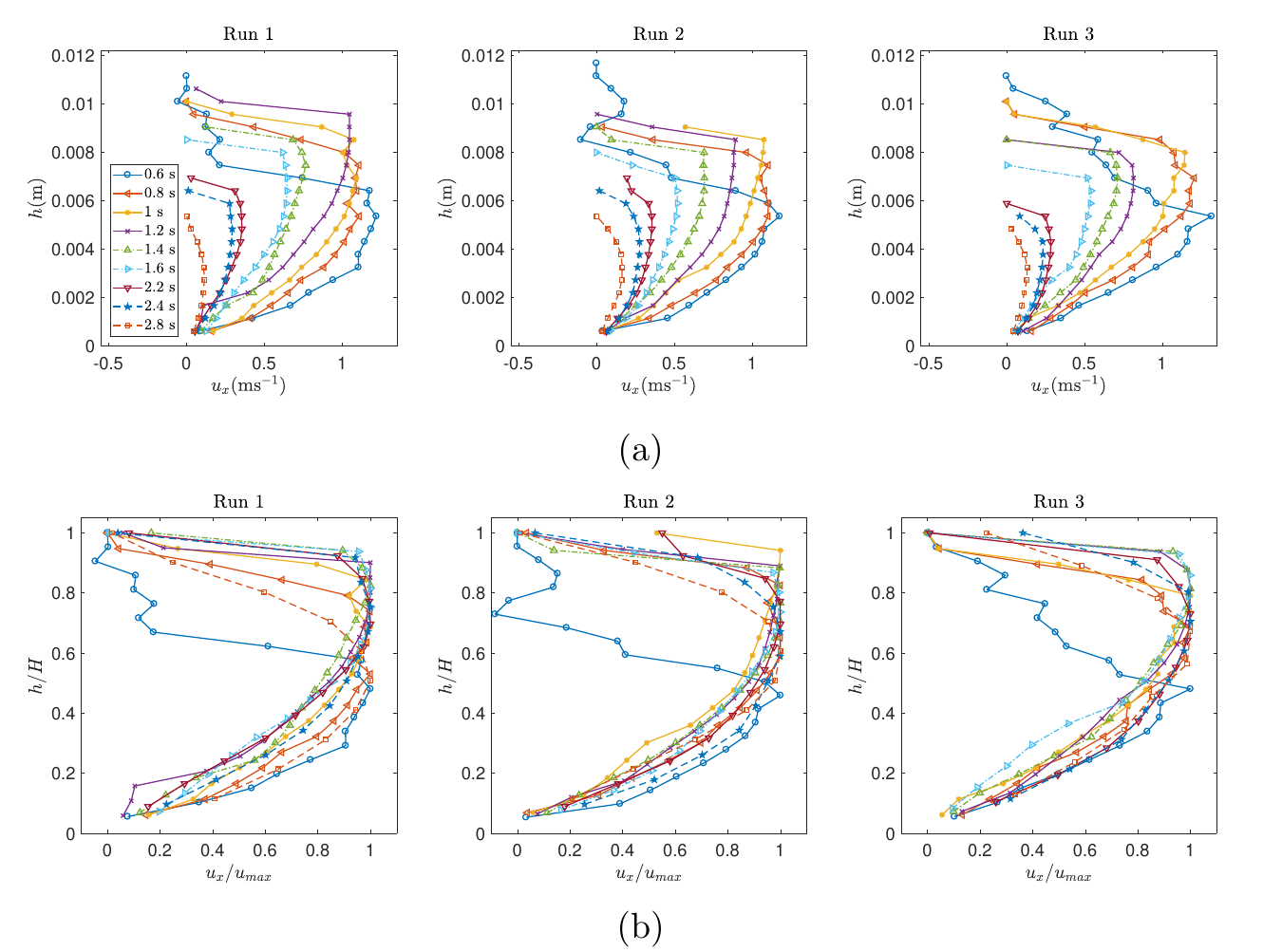}%
\caption{Plots of a) horizontal velocity, and b) normalised horizontal velocity, against height at $x = 1.102$ m downstream from the lock gate. On each graph the plotted profiles are from eight different flow times, ranging between $t = 0.6$ s and $t = 2.8$ s, for the three different experimental runs.}
\label{fig13}
\end{figure}

We now neglect the initial purely collisional regions of the flow, and focus our attention on the behaviour from $t = 0.6$ s onward. The velocity profiles at different times from $t = 0.6$ s to $t = 2.8$ s are plotted together in Figure \ref{fig13}a, for each experimental run. It can be seen that the velocities are highest at $t = 0.6$ s, where the maximum value is approximately $u_x = 1.2$ m s$^{-1}$. The maximum velocities, as well as the velocities below the height of the maximum, decrease with time. The height of the velocity maximum increases from $t = 0.6$ s to $t = 1.2$ s, where it then decreases until $t = 2.8$ s. The velocity profiles were normalised by dividing the velocity and the height by the maximum velocity $u_{max}$ and the flow depth $H$ respectively. Profiles of normalised velocity are plotted against normalised height in Figure \ref{fig13}b. For all runs, the majority of the normalised profiles approximately collapse onto one curve. A common exception is the profile at $t = 0.6$ s, which has a normalised height of the velocity maximum that is significantly lower than for the later times. Furthermore, there is some variation between the profiles at $t = 0.8$ s, $t = 1$ s and $t = 2.8$ s. The three runs are consistent in that the velocity profiles between $t = 1.2$ and $t = 2.4$ are almost indistinguishable from one another. At these times, the velocity increases linearly with height in the region above the flume bed, from a value of approximately zero (at the bed). The velocity gradient increases as the height approaches that of the velocity maximum, which has a non-dimensional height of approximately $0.8$. Above this height, the velocity decreases towards the surface, although the velocity data in this area is subject to error, as discussed above. The collapse of the velocity profiles onto a single curve has been observed in previous experimental investigations of debris flows \citep{kaitna2014surface,sanvitale2016visualization}, and indicates the uniform behaviour within this region of the flow. 

\subsection{\textit{Shear behaviour}}

\begin{figure}
\centering
\includegraphics[width=0.9\textwidth]{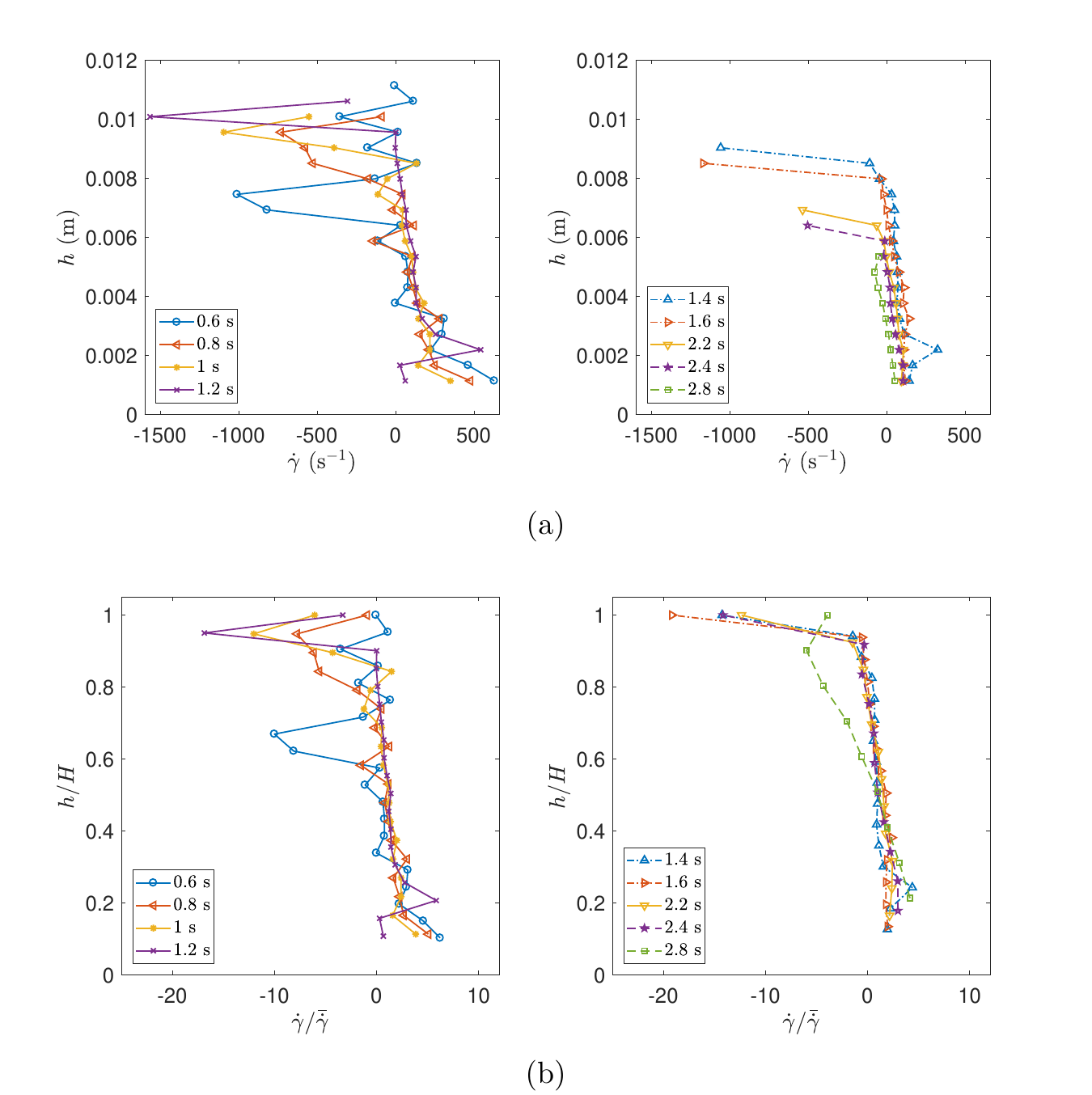}%
\caption{Plots of a) shear stress rate, and b) normalised shear stress rate, against height at $x = 1.102$ m downstream from the lock gate, for Run 1 of the experiment.}
\label{fig14}
\end{figure}

The PIV data can be utilised to obtain profiles of the internal shear strain rate.
Neglecting the horizontal gradients of the vertical velocity $u_y$, the local shear strain rate $\dot{\gamma}$ is defined as 
\begin{equation}
\dot{\gamma} = \frac{\partial u_x}{\partial y}.
\end{equation}
The shear rate is approximated at each vertical height $h_i$:
\begin{equation}
\dot{\gamma}_i \approx \frac{u_{x,i+1} - u_{x,i}}{\Delta h}, 
\end{equation}
where $u_{x,i}$ is the velocity at the current vertical position $h_i$, and $u_{x,i+1}$ is the velocity at the subsequent vertical location $h_{i+1}$. The vertical sampling points are separated by the distance $\Delta h$. The profiles of local shear rate plotted against height are shown in Figure \ref{fig14}a for one of the experimental runs. In order to visualise the results more clearly, the profiles from $t = 0.6$ s to $t = 1.2$ s are plotted separately from the results between $t=1.4$ s and $t = 2.8$ s.
From $t = 0.6$ to $t = 1$ s, the shear is highest at the bed and sharply decreases in the region just above the bed. Above this, the overall shear decreases as the height increases, although the values fluctuate locally. At the boundary between the non-fluctuating and collisional regions of the flow at $t = 0.6$ s and $t = 0.8$ s (see Figures \ref{fig8} and \ref{fig9}), the shear rate decreases to a large negative value. This is followed by an increase in shear above the interface between the two regions, highlighting the presence of a shearing layer in this region. The shear rate profiles at $t = 1$ s and $t = 1.2$ s display similar behaviour in that there is a steep increase in the region directly above the bed, followed by a uniform decrease until just below the free surface. At the free surface, there is a sudden decrease to a negative shear value. The profiles of internal shear rate between $t = 1.4$ s and $2.4$ s exhibit a linear decrease with height from the flume bed, until the free surface. Again, in this region the shear profile sharply decreases to a large negative value. The spurious negative shear values at the free surface reflect the relatively large deviations from the velocity average in the area (see Figure \ref{fig8}). At the final considered time of $t = 2.8$ s, the shear rate profile decreases linearly until just below the free surface, before increasing towards the surface.    

Following \citet{sanvitale2016visualization}, the normalised shear rate $\hat{\dot{\gamma}}$ is obtained by dividing by the depth-averaged shear rate $\bar{\dot{\gamma}}$:
\begin{equation}
\hat{\dot{\gamma}} = \frac{\dot{\gamma}}{\bar{\dot{\gamma}}} = \frac{\partial u_x}{\partial y} \frac{H}{(u_H - u_{slip})},
\label{eq:normal_shear}
\end{equation}
where $u_H$ and $u_{slip}$ are the values of horizontal velocity at the free surface and the bed respectively. Note that if both $u_H$ and $u_{slip}$ are zero, Equation \eqref{eq:normal_shear} is inapplicable. In the current results,  $u_H$ and $u_{slip}$ are small in comparison to the internal velocities, yet have non-zero values. Profiles of normalised local shear $\hat{\dot{\gamma}}$ are provided in Figure \ref{fig14}b. Bar a difference near the flume bed, the shear rate profiles from $t = 1.4$ s to $t = 2.4$ s almost collapse onto the same profile.

\section{Discussion}
\label{sec:exp_discussion}

\subsection{\textit{Overall spatial and temporal flow evolution}}

The experimental debris flow described in the previous section exhibits complex behaviour throughout its evolution. This is immediately evident from the snapshots shown in Figure \ref{fig7}. The front of the flow is dilute and turbulent, while the flow body develops into what appears to be a steady, water-granular mixture. In the transition between these two types of behaviour, the flow is composed of two distinct layers. The data obtained from the PIV analysis of the flow has allowed further insight into the mechanisms that are responsible for the observed experimental behaviour. By defining a threshold value of deviation from the locally averaged velocity, it is possible to distinguish two types of flow behaviour -- non-fluctuating and collisional \citep{paleo2014fluid}. In the current analysis, the threshold between the two types of flow has been defined as a standard deviation that is $20 \%$ of the local average velocity. Furthermore, a constant threshold of $\bar{e} = 0.15$ m s$^{-1}$ has also been implemented, following \citet{paleo2014fluid}. In both cases, contour plots of the velocity deviation (see Figures \ref{fig8} and \ref{fig9}) have shown that the dilute front of the flow is dominated by collisional and turbulent behaviour. Behind the front, the flow transitions from collisional to non-fluctuating, and consists of both types of behaviour in between. Throughout the transition, the collisional region (with a high standard deviation) narrows in depth, decreases in magnitude and shifts towards the flow free surface. Simultaneously, the height of the non-fluctuating region increases with distance from the bed, until the flow is dominated by this type of behaviour. The non-fluctuating region corresponds to a shearing layer, where the mixture shears along the bed. This can be deduced from the velocity contours and vertical profiles (see Figures \ref{fig10} and \ref{fig11}), where the initial presence of the non-fluctuating region coincides with the initiation of a layer of shear between the mixture and the bed. The height of the shearing layer increases due to the accumulation of granular material that is opposed by the frictional resistance of the bed, until the entire flow consists of a steady shear flow.

In the field, subaerial debris flows are typically composed of a dry head containing large particles, where the dynamics are dominated by granular collisions and exhibit frictional behaviour. Behind the head, the flow body contains smaller particles and interstitial fluid, exhibiting fluid-like behaviour \citep{iverson1997physics,mcardell2007field}. The head-body architecture is attributed to complex couplings related to the grain size distribution, material fines content and pore water pressures \citep{iverson2010perfect,johnson2012grain}. It has also been observed in small scale experiments \citep{parsons2001experimental,kaitna2014surface,sanvitale2016visualization}, including those with monodispersed spherical mixtures \citep{paleo2014fluid}. In the current work, the high water content of the experimental mixture does not permit the formation of a dry, granular head, yet the front of the flow is also dominated by granular collisions, along with turbulent fluid behaviour. Furthermore, the flow transitions from the collisional front to a distinct body, characterised by non-fluctuating behaviour (see Figures \ref{fig8} and \ref{fig9}). The experiments therefore exhibit a unique head-body transition, while still retaining similarities to classic debris flows. This difference may be important when applying models derived from subaerial debris flows to subaqueous debris flows (e.g. \cite{felix2006transformation}, \cite{felix2009field}, \cite{ducassou2013run}, \cite{paull2018powerful}). 

Regarding small scale debris flow experiments within the literature, internal flow observations have contributed to a better understanding of the dynamics responsible for debris flow architecture \citep{kaitna2016effects}. The majority of attention has previously been focused on the internal dynamics of steady granular flows, in order to approximate the rheology of granular material \citep{armanini2005rheological,kaitna2007experimental,kaitna2014surface}. Furthermore, in experiments of unsteady debris flows, the internal behaviour of the collisional flow head has not been taken into consideration. This has been a result of restrictions in software \citep{sanvitale2016visualization}, or because the high velocity fluctuations in this region are the source of unreliable data \citep{paleo2014fluid}. In the experiments of \citet{sanvitale2016visualization}, the flow head was dry -- the laser-based PIV technique can only detect saturated particles. In the present analysis, the internal velocity profile has been examined within both the front of the flow and the body, which has highlighted how the flow progressively transitions between the collisional and non-fluctuating regimes. One of the most striking features of this transition is the presence of two distinct shearing layers, which can be seen at $t = 0.6$ s and $t = 0.8$ s (see Figures \ref{fig8}, \ref{fig10} and \ref{fig11}). It should also be highlighted that the majority of previous investigations into internal granular flow dynamics have presented profiles of velocity, and the temporal evolution of the depth-averaged velocity \citep{armanini2005rheological,kaitna2014surface,paleo2014fluid,sanvitale2016visualization}. In addition to velocity profiles, we have also produced velocity flow fields, which clearly present the initiation and evolution of a granular shear layer (see Figure \ref{fig11}). 

\subsection{\textit{Velocity profiles within the flow body}}

\subsubsection*{\textit{Solid bed flow}}

\begin{figure}
\centering
\includegraphics{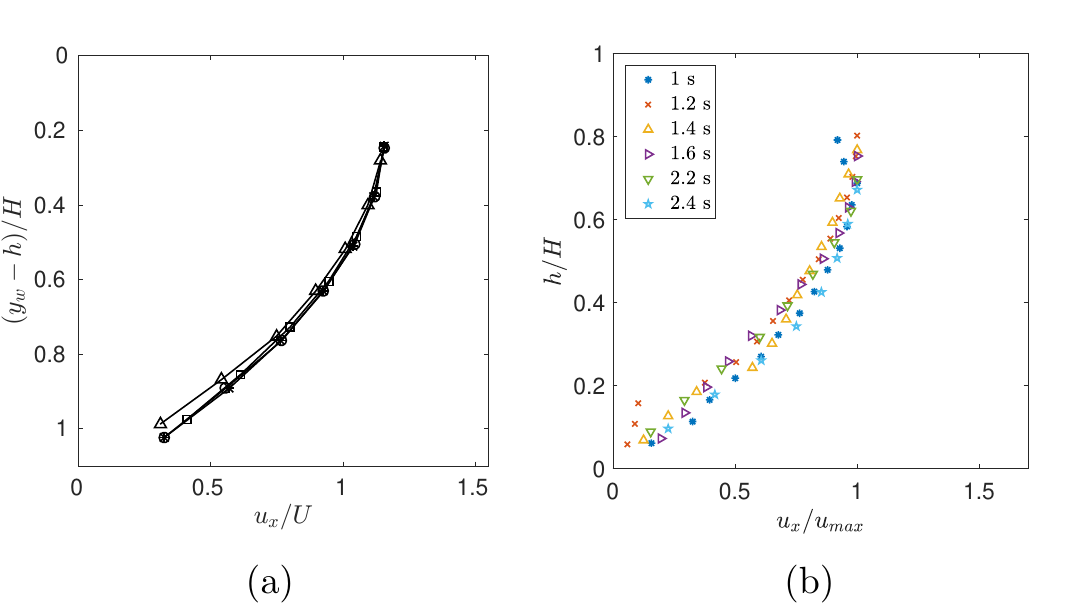}
\caption{The normalised internal velocity profiles in the steady, solid bed flow experiments of a) \citet{armanini2005rheological} (adapted from \citet{armanini2005rheological}), compared with b) the velocity profiles in the body of the current experiment (Run 1). For the former, plotted on the $x$-axis is horizontal velocity normalised by the mean velocity $U$. The $y$-axis shows $(y_w-h)/H$, where $y_w$ is the saturation line (obtained by visual inspection). The points correspond to the experimental velocity values, where the different symbols refer to results from different runs with the bed slope varying from $19^{\circ} - 23 ^{\circ}$.}
\label{fig15}
\end{figure}

Considering the behaviour within the body of the experiment, the internal structure can be compared against that of existing experimental debris flows. The relevant details of a selection of experiments presented in the literature are provided in Table \ref{tab:lit_exp}, along with the parameters in the current debris flow experiments. Disregarding the spurious values at the free surface (the upper four data points --- corresponding to 1.6 mm), the profiles in the present investigation are similar to those of the solid bed flow described by \citet{armanini2005rheological}, from $t = 1$ s onwards. Solid bed flows occurred for the highest bed inclinations, and were characterised by the shearing flow of the granular phase over a fixed bed. Profiles of horizontal velocity exhibited a convex shape, that increased with distance from the bed. The corresponding shear rate profiles decreased with distance from the bed, to a near zero value. The solid bed velocity profiles of \citet{armanini2005rheological} are compared against those of the current work in Figure \ref{fig15} --- the profiles clearly exhibit the same behaviour. Similar convex velocity profiles have also been recorded in the body of a steady debris flow consisting of gravel and water, in a rotating drum \citep{kaitna2014surface}. The experiments were conducted for four different mixtures of gravel, mud and water, and it was found that each mixture exhibited a distinct, dimensionless velocity profile, for a number of different drum rotation velocities. Furthermore, solid bed profiles have been observed in the body of an unsteady, experimental debris flow, that was similar in set-up to that of the current work \citep{sanvitale2016visualization}. The normalised velocity profiles at different flow times approximately collapsed onto a single curve. Similar to the findings of the \citet{kaitna2014surface}, a distinct profile was exhibited for three different grain size distributions, mixed with water. In the current work, the normalised velocity profiles in the flow body also collapse onto a single curve for each experimental run, from $t = 1$ s onwards (see Figure \ref{fig15}b), showing the flow similarity over this time. 

\begin{landscape}
\begin{table}
\caption{A summary of the relevant experimental parameters in the current work, and selected investigations in the literature. The notation $x_f$ refers to the flume run-out length (from the lock gate position), $c_w$ is the channel width, $d_d$ is the diameter of the rotating drum, $\theta$ is the flume inclination, $\phi_s$ is solid volume fraction, $d_{50}$ is the mean particle size and $C_U$ is the coefficient of uniformity. The abbreviation n.p. denotes information that was not provided in the literature.}
\centering
\scriptsize
\begin{tabular}{c|c c c c c c c} 
Reference & Flume dimensions (m) & Inclination $\theta$ ($^{\circ}$) &  Volume fraction $\phi_s$ &  Sediment &  Fluid &  $d_{50}$ (mm) &  $C_U$ \\
\hline
 Present work &  $x_f = 1.662$, $c_w = 0.1$ &  31 & 0.44 &  Crushed glass grit &  Water &  0.917 & 5 \\
\citet{armanini2005rheological} &  $x_f = 6$, $c_w = 0.2$ & $5 \leq \theta \leq 22$ & 0.346 - 0.529  &  PVC pellets & Water & 3.7 & 1 \\
\citet{kaitna2014surface} & $d_d = 4$, $c_w = 0.8$ & n/a & 0.6, 0.62, 0.7 &  Gravel & Water, mud  & 4, 8, 10, 13 & n.p. \\
\citet{paleo2014fluid} & $x_f = 0.7$, $c_w = 0.156$ & 27 & 0.4 &  Glass spheres & Water, glycerol & 2 & 1 \\
\citet{sanvitale2016visualization} & $x_f = 2$, $c_w = 0.15$ &  24.5 & 0.7 & Crushed and subrounded glass & Hydrocarbon fluid & 7.1 & 3,10,20 
\end{tabular}
\label{tab:lit_exp}
\end{table}
\end{landscape}

\subsubsection*{\textit{Viscous and granular scaling}}

\begin{figure}
\centering
\includegraphics[width=0.7\textwidth]{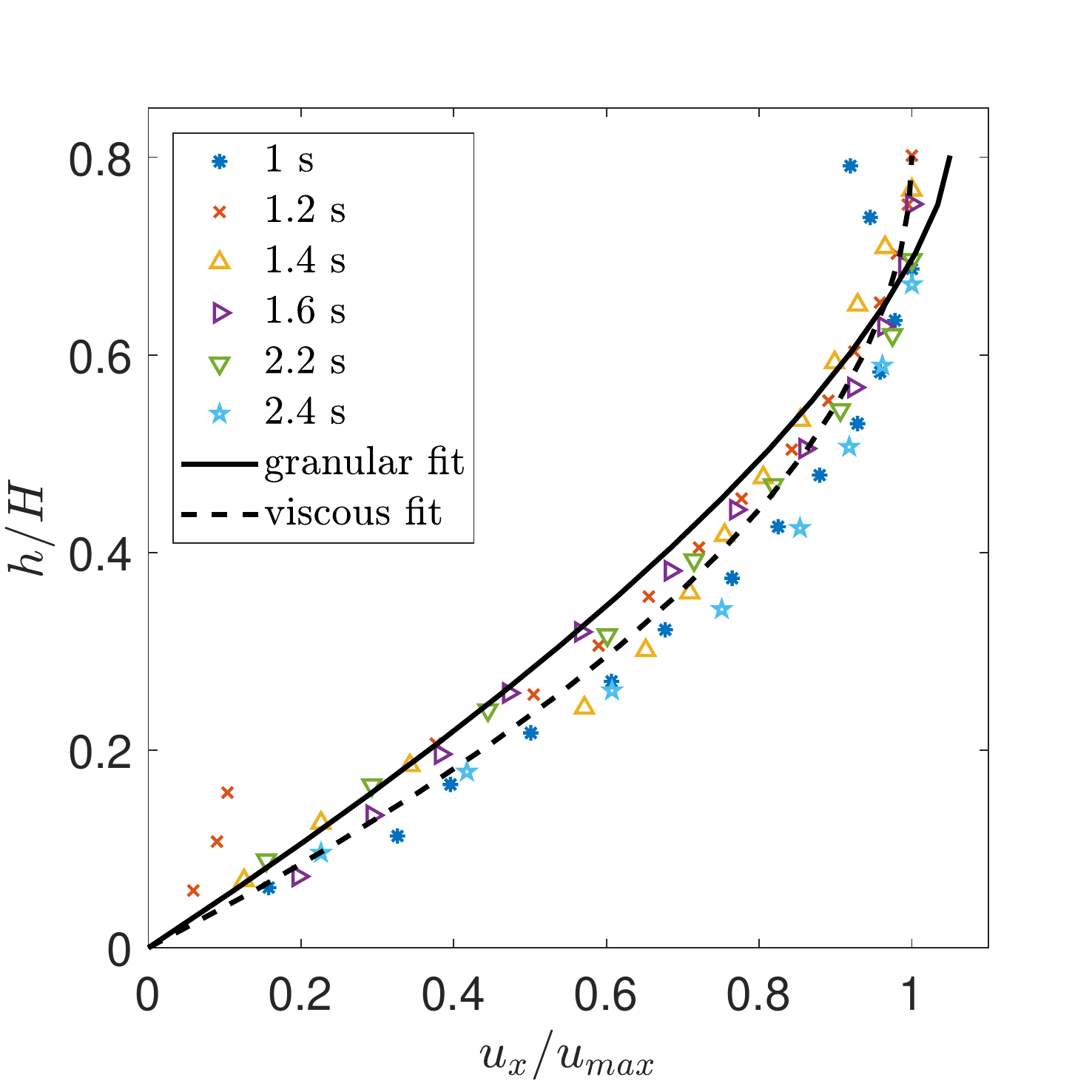}
\caption{Normalised velocity profiles in the sheared region for Run 1 of the experimental debris flow, with a best fit granular and viscous scaling.}
\label{fig16}
\end{figure}

\citet{bagnoldexperiments} examined the rheological behaviour of sheared mixtures of suspended, non-cohesive spheres. Two regimes were identified --- viscous and granular --- which were distinguished according to a dimensionless number describing the ratio of internal grain stresses to fluid stresses. Applying Bagnold's findings to a uniform, steady flow results in two theoretical vertical velocity profiles. For a viscous-type flow, the velocity profile scales as $u_x(y) \propto H^2 - (H-y)^2$, where $H$ is the height of the free surface. Alternatively, the velocity in a granular regime (dominated by granular collisions) scales as $u(y) \propto H^{\frac{3}{2}} - (H-y)^{\frac{3}{2}}$. We assess the rheological behaviour in the experimental flow body by approximating the dimensionless velocity profile with a granular and viscous scaling. Figure \ref{fig16} shows the normalised velocity profiles for Run 1 of the experiments between times of $t = 1$ s to $t = 2.4$ s, where the upper $1.6$ mm of flow has been omitted. The profile of best fit has been included, assuming both a viscous and granular scaling. The closest fit to the experimental results is found with the viscous scaling, which captures the overall velocity shape throughout the shear layer. In the experiments of \citet{sanvitale2016visualization}, mixtures with a wide grain size distribution (with a coefficient of uniformity of $C_U = 20$) exhibited a viscous-type velocity profile. Conversely, for $C_U = 3$ a granular profile provided the best fit to the experimental data. A wider grain size distribution promotes particle segregation, which can lead to the finer particles being trapped within the solid matrix. The presence of these fine grains enhances the viscosity of the interstitial fluid, and viscous forces influence flow behaviour \citep{iverson1997physics}. For a more uniform particle distribution, the dominating forces are generally inter-particle granular collisions. In the current experiments, the viscous profile provides the closest fit to the experimental results, despite a relatively small coefficient of uniformity of $C_U = 5$. This is possibly due to the proportion of very small particles with diameters less than 0.5 mm that are present within the mixture (see Figure \ref{fig2}), which add to the fluid viscosity. The results also suggest that the cut-off between granular and viscous-type flow may lie between $C_U = 3$ and $C_U = 5$. A suggested area for future work is to perform further experiments with different values of $C_U$, to test this hypothesis.

\subsection{\textit{Fluid-particulate coupling}}

The current experiments were performed with a significantly higher content of fluid than for the majority of similar, debris flow experiments (see Table \ref{tab:lit_exp}). Despite this, the behaviour observed within the flow body is comparable to results presented in the literature, as discussed above. In terms of volume fraction, one set of a series of flume experiments performed by \citet{paleo2014fluid} used a value of $\phi_s = 0.4$ (which is close to $\phi_s = 0.44$ in the current work). The mixture in their work consisted of water and monosized glass spheres with a diameter of 2 mm. The height evolution of the two phases was recorded at a distance $0.232$ m downstream from the lock gate. Upon arrival at this location, the flow exhibited a dry, granular front. The height of the water phase increased with time, until the flow body was composed of a water-granular mixture, with an upper layer of water. The entire solid phase had propagated beyond the location of measurement by approximately 1.2 seconds, and the tail of the mixture consisted of water only. Conversely, for a higher solid volume fraction of $\phi_s = 0.6$, the tail of the flow contained a mixture of both spheres and water of approximately the same height (although a larger section of the flow front was dry). This behaviour suggests that for the higher water content ($\phi_s = 0.4$), the coupling between the two phases was less significant than in the current work. The current experimental flow did not exhibit any regions of dry granular material and had a tail composed of both water and granular material. The strong difference in behaviour between our experiments and those of \citet{paleo2014fluid} is attributed to the composition of the granular phase. As opposed to monosized spheres, the current experiment consists of crushed glass of varying diameter. The angular shape allows the interlocking of grains and adds extra frictional resistance that is not present for spherical grains. For angular, crushed material, inter-particle shearing is significant, in addition to the shearing between the material and the bed. Therefore a flow consisting of realistic granular material exhibits lower velocities than that of glass spheres. Furthermore, the dilation and contraction of the crushed glass particles regulates the motion of the water, enhancing the coupling between the two phases. Experiments involving spherical particles are beneficial in terms of simplicity, and allow the investigation of a wide range of factors affecting flow behaviour. However, the flow dynamics can be significantly different from that of a realistic granular material, as shown by the qualitative difference between the current experimental results and those of \citet{paleo2014fluid}. The relatively high water content in these experiments has particular relevance to subaqueous debris flows, which have wet heads. Based on the present experiments, subaqueous debris flows may exhibit markedly different flow behaviour at the head than current models (derived from subearial debris flows) predict.

\subsection{\textit{Velocity profiles within the head-body transitional region}}

In the transitional region between the head and body of the experimental debris flow, it is comprised of a concentrated lower layer and a more dilute upper layer (see $t=0.3$ s and $t = 0.6$ s in Figure \ref{fig7}). The corresponding velocity profiles show that in the lower layer, the velocity increases with distance from the bed to a velocity maximum towards the top of the layer. Above the maximum, the velocity decreases rapidly and exhibits negative values due to shearing between the layers, and the inability of the PIV software to produce accurate velocity values. These profiles share similarities with those observed in the steady state profiles of some submarine gravity currents, where differences in density drive a dense fluid through a less dense, ambient fluid \citep{simpson1979dynamics,kneller1999velocity,lowe2002laboratory}. 
For some sediment-laden flows, notably high concentration turbidity currents, the settling of sediment can result in a layer of high concentration at the bed, while the upward mixing of turbulence produces a dilute upper layer that entrains sediment \citep{postma1988large,stevenson2018reconstructing}. In the internal profiles of these flows, the velocity maximum is located at the top of the lower layer due to the balance of the shear at the bed and at the interface between the dense fluid and the ambient fluid \citep{kneller1999velocity}. These profiles are observed in steady state flows, and above the interface between the two layers of material the velocity steadily approaches a zero value. This overall shape is similar to the internal velocity profiles in the current experimental flow at $t = 0.3$ s and $t = 0.6$ s (see Figure \ref{fig12}). Although the flow is transient at these times, and shows large fluctuations in the upper layer, the analogy to high concentration turbidity currents provides a deeper understanding of the mechanism responsible for the observed velocity profiles. Furthermore, it has been postulated that the transport of sediment in high concentration turbidity currents is a result of the interaction between a high concentration lower layer, and a turbulent upper layer \citep{postma1988large}. This has potential relevance to the formation of the observed architecture in the current flow. However, it should be noted that two-layer turbidity currents are only a subset of natural systems (e.g. \citet{paull2018powerful}), and many flows are likely to exhibit a more gradual stratification \citep{peakall2015submarine}. 

\subsection{\textit{Experimental limitations}}

As discussed in Section \ref{sec:piv}, the PIV method requires the detection of individual particles over multiple frames in order to produce accurate velocity vectors. This wasn't possible at the head of the flow due to the low particle concentration and their turbulent behaviour. The velocity values recorded at the flow free surface were also subject to error, due to an overlying layer of water where particles were not present. The fact that particle tracking was not accurate at the front of the flow suggests that some particles were transported away from the flume walls in the cross stream direction, as a result of the high fluid turbulence. This implies that the two-dimensional flow approximation is subject to error, particularly at the front of the flow. Furthermore, the presence of the side wall may influence the flow dynamics. Despite these limitations, the experiments showed a high degree of repeatability, as shown in Figure \ref{fig12}. The small differences between the different runs at certain times may be a result of a delay in the opening of the lock gate, or variability in material composition.

\section{Conclusion}
\label{sec:conc}

The internal observations of the experimental debris flow have provided insight into the complex interaction between propagating particulate and water phases. The experiments consisted of a relatively dilute water-granular mixture ($\phi_s = 0.44$), which exhibited a spatial and temporal evolution from a transient, collisional, turbulent flow to a steady, non-fluctuating flow. The transition from the collisional flow front to the non-fluctuating body has been quantified in a small scale unsteady debris flow for the first time. A head-body architecture was observed during the flow evolution, the type of which has not been documented before. The initially turbulent head evolved into a uniform flow as a layer of shearing granular material increased in height. Unlike similar experiments with monosized spheres \citep{paleo2014fluid}, the body of the current flow exhibited a thin layer of water overlying the viscous mixture for the entirety of the flow duration. This indicates that the behaviour of the flow was influenced strongly by the coupling of the granular and liquid constituents. Indeed, in reality granular-fluid coupling plays a vital role in debris flow dynamics \citep{iverson2003debris,iverson2011positive}. The experimental flow in the current work is therefore more analogous to realistic debris flows than experiments involving idealised spherical particles, and the data could provide valuable validation for the development of two-phase numerical models. Regarding the non-fluctuating flow body, a viscous-type profile is able to capture the velocity throughout the majority of its depth. The granular material in the current experiments has a coefficient of uniformity $C_U$ of 5. Integration with the work of \citet{sanvitale2016visualization} suggests that the transition from a granular to a viscous-type flow profile takes place between a coefficient of uniformity of 3 and 5.

\section*{Acknowledgements}

This work was supported by the Engineering and Physical Sciences Research Council (EPSRC) Centre for Doctoral Training in Fluid Dynamics at the University of Leeds under grant no. EP/L01615X/1. The experiments were conducted in the NERC recognized Sorby Environmental Fluid Dynamics Laboratory at the University of Leeds. We thank Helena Brown and Rob Thomas for their invaluable help with the execution of the experiments.

\bibliography{main_compressed}

%\begin{thebibliography}{}
%\bibitem[Peterman(1982)]{peterman1982}
%Peterman, R.M. 1982. Model of salmon age structure and its use in  preseason forecasting and studies of marine  survival. Can.~J.~Fish.~Aquat.~Sci.
%\textbf{39}: 1444--1452.
%\end{thebibliography}

\end{document}